\documentclass[10pt]{article}
\usepackage{apacite}
\usepackage[utf8]{inputenc}
\usepackage{amsmath,amsfonts,amsthm,amssymb}
\usepackage{tabularx}
\usepackage{multirow}
\usepackage{graphicx}
\usepackage{natbib}
\usepackage{ragged2e} 
\usepackage{float}
\usepackage{subcaption}
\usepackage[font={small,bf}]{caption}
\usepackage[margin=1in]{geometry}
\usepackage{chngcntr} % Number figures based on section
\counterwithin{figure}{section}
\counterwithin{equation}{section}
\counterwithin{table}{section}
\title{System Identification of the Upgraded LHPOST6 Reaction Mass at the University of California San Diego}
\author{Andres Rodriguez-Burneo$^1$, Jose I. Restrepo$^1$, Joel P. Conte$^1$ }
\date{$^1$University of California San Diego, Department of Structural Engineering}
\begin{document}
%%%%%%%%%%%%%%%%%%%%%%%%%%%

%%%%%%%%%%%%%%%%%%%%%%%%%%%
\maketitle
%%%%%%%%%%%%%%%%%%%%%%%%%%%

%%%%%%%%%%%%%%%%%%%%%%%%%%%
\begin{abstract}
%%%%%%%%%%%%%%%%%%%%%%%%%%%

\justify
\it{Upon completing the upgrade from one to six degrees of freedom of the Outdoor Shake Table at UCSD in 2019, forced vibration tests were carried out to identify the dynamic characteristics of the reaction mass and soil system. This report describes the motivation, execution, and results from such tests, which independently excited the reaction mass in four degrees of freedom: longitudinal, transverse, yaw, and vertical. The report discusses the frequency response curves and deformation patterns from which the natural frequencies, damping ratio, mode shapes, and rigid body motion were determined. The first objective of the study was to investigate if the dynamic properties of the system had dramatically changed after the upgrade by comparing the results to those from forced vibration tests performed 20 years ago, during the construction of the facility. In addition, most recent tests also contributed with results from the vertical degree of freedom, which had never been tested. The second objective was to obtain high-quality response data of the system that will be used to develop a high-fidelity computational model of the reaction mass in future research. A comparison of results showed a slight difference of 0.5Hz in the natural frequency of 2 degrees of freedom. Moreover, maximum displacements in the recent tests were overall larger than the previous ones with few exceptions. The report thoroughly discusses the several sources of discrepancy between the past and most recent results. Finally, test results allowed us to estimate the system's response if the shake table actuators were to be used at their maximum nominal capacity. Small displacement and high damping results were consistent with those of previous tests and further validated the design of the reaction mass.\\

}
\end{abstract}

\newpage
\tableofcontents

\newpage
\listoffigures

\newpage
\listoftables

\newpage
%%%%%%%%%%%%%%%%%%%%%%%%%%%
\section{INTRODUCTION}
%%%%%%%%%%%%%%%%%%%%%%%%%%%

In October 2002, the University of California San Diego was commissioned by the national Science Foundation (NSF) to build a Large High-Performance Outdoor shake table (LHPOST). This facility was designed to be part of the Network for Earthquake Engineering Simulation (NEES) program. From its conception, the LHPOST was designed to be the first outdoor and largest shake table in the United States, capable of reproducing far and near field ground motions on full-scale specimens \citep{VDE:2004}. 

From the beginning, the LHPOST, composed of the shake table system and the surrounding foundation block or reaction mass, was conceived to operate with six-degrees-of-freedom (6-DOF) capabilities. However, its development was divided into two phases, where the first one allowed the operation of the LHPOST only in a single degree of freedom \citep{Restrepo:2005}. During the first phase of construction, forced vibration studies were performed to obtain dynamic data of the foundation-soil system and understand the effects of the vibration of the shake table on the reaction mass. Results from such study validated the design of the reaction mass, capable of withstanding the forces applied by the actuators, and provided invaluable data for the future development of a virtual model of the shake table facility, including soil-structure interaction (SSI). \citep{Luco:2011}  

The LHPOST operated in the 1-DOF configuration between 2004 and 2019, when works on the 6-DOF upgrade began. Over 30 major projects were conducted on the LHPOST in full and distorted scales during this time. Some examples include fixed base and base-isolated buildings, steel-framed, reinforced concrete, timber, masonry structures, and infrastructure such as wind turbines, bridge piers, and soil retaining walls. In addition to the shake table, the complex features a blast/impact facility and a soil-structure interaction facility, depicted in Figure 1.1. Van Den Einde et al. (\citeyear{VDE:2021}) describes an exhaustive list of the projects performed. The design of the upgraded shake table was a collaboration between UCSD and MTS System corporation. The 6-DOF upgrade required modifications on all the main components of the LHPOSTS including the actuators, hydraulic system, and surrounding reaction mass, depicted in Figure 1.2

The steel platen has a footprint of 12.2m x 7.6m with a maximum vertical payload of 20MN, making it the largest 6-DOF outdoor shake table in the world with the largest payload capacity. The surrounding reaction mass has overall dimensions of 33.12m in length and 16.91m wide, extending to a depth of 5.79m with a smaller area in the center of the foundation reaching a depth of 7.92m. In general, the weight of a reaction mass is expected to be around 20 times larger than the maximum payload capacity, to guarantee a low natural frequency far away from those of the specimens being tested. However, the LHPOST reaction mass is only 2.2 times heavier than the upgraded payload capacity. The relatively low mass and unconventional geometry of the reaction mass aimed to take advantage of the stiff soil to maximize radiation damping, controlling the amplitude of the response. This approach not only accounted for changes foreseen in the upgrade, but also resulted in a smaller, lighter, and more economic design \citep{Restrepo:2005}.

\begin{figure}[H]
 \centering
 {\includegraphics[width=.55\linewidth]{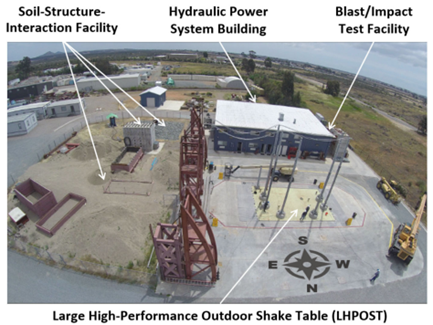} }
 \caption{View of the entire facility \citep{VDE:2021}}
  \label{fig1.1}
\end{figure}

To achieve the 6-DOF capabilities. The new configuration doubled the number of horizontal actuators, re-arranged them in a V-shape and added 6 new vertical actuators underneath the platen. The combined force of the horizontal actuators is limited to 6.8MN and 3.4MN in the East-West and North-South directions respectively.  The vertical actuators were installed in 2009, acting only as pressure balance bearings but could not move the platen vertically until the upgrade was completed. Moreover, given the new displacement capabilities, a system with three lines of defense was designed to protect against the impact of the platen onto the reaction mass. The protection system includes software limit detectors, physical limit switches (LVDT's), and the crash protection bumpers shown on Figure 1.2. In addition, the hydraulic power system was improved to supply the new actuator demands. The system included the accumulator banks, blow-down valves, pumps, and pressure lines. Finally, removable concrete covers, and steel plate covers, were re-designed to protect the actuators from falling debris without compromising the movement of the platen or the actuators \citep{VDE:2021}. 

With the upgraded force and displacement capabilities of the shake table, a new forced vibration study was performed to evaluate the dynamic response of the reaction mass-soil system. The first objective of the study was to determine if the properties of the system had dramatically changed after the upgrade. The second was to obtain high-quality response data of the system that will be used to develop a high-fidelity computational model of the reaction mass in future research, out of the scope of this report.   

\begin{figure}[H]
 \centering
 {\includegraphics[width=.65\linewidth]{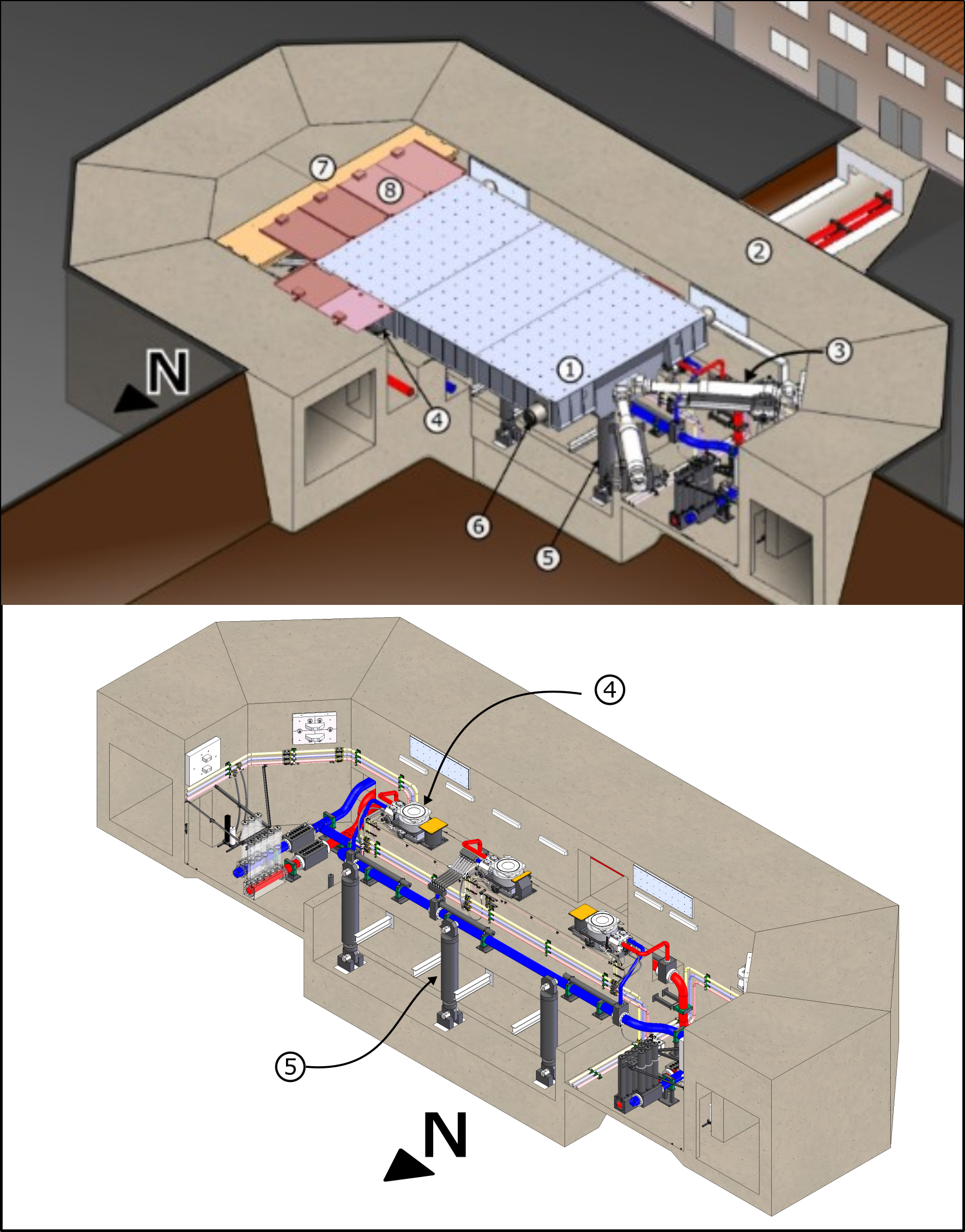} }
 \caption{Components of the upgrades 6-DOF LHPOST Facility. Description of the facility and 6-DOF upgrade. (1) Steel Platen, (2) reaction mass, (3) horizontal actuators, (4) vertical actuators, (5) hold-down-struts, (6) crash protection bumpers, (7) upgraded concrete cover (8) steel plates}
  \label{fig1.2}
\end{figure}

\subsection{Soil Characteristics}

Several field investigations have been performed to characterize the soil since the initiation of the project. Results from 2002 showed three main geologic units surrounding the site: top soils characterized as firm, sandy clay with gravel and cobble and loose clayey fine to very coarse sand with gravel and cobble covering the site to a depth between 0.6m-0.9m. Linda vista geologic formation extend further down to approximately 3.7m. Which is constituted by very dense, clayey, fine to very coarse sand with gravel and cobble. Tertiary soils of the stadium conglomerate geologic formation were found underneath. These are constituted by very dense, silty, fine to very coarse sand to sandy, cobbly gravel. The maximum depth of the borehole investigation reached 22 m and did not encounter the water table. 

Furthermore, top soils and stadium conglomerate presented a dry unit weight of 16.8 kN/m$^3$ and 18 kN/m$^3$, respectively obtained from laboratory tests. Moreover, the soil bearing capacity was determined as 191kPa at the surface plus 157 kPa per additional meter of depth with a maximum of 479 kPa. Finally, shear wave velocity form tests carried out in 2012 resulted in values between 185m/s - 305m/s and 760m/s for the Linda vista formation and stadium conglomerate, respectively \citep{Luco:2011}.  The current soil profile surrounding the reaction mass is depicted in Figure 1.3.

After completion of the first phase of development, backfill was placed surrounding the sides of the reaction mass and compacted to 90\%  of the proctor standard, leaving the stadium conglomerate in contact only with the bottom of the structure. An important factor worth mentioning is the possibility of replacing the soil in the SSI facility at the east of the shake table, and modifying the characteristics of the surrounding soil in the vicinity of the reaction mass, see Figure 1.1. 

\begin{figure}[H]
 \centering
 {\includegraphics[width=.51\linewidth]{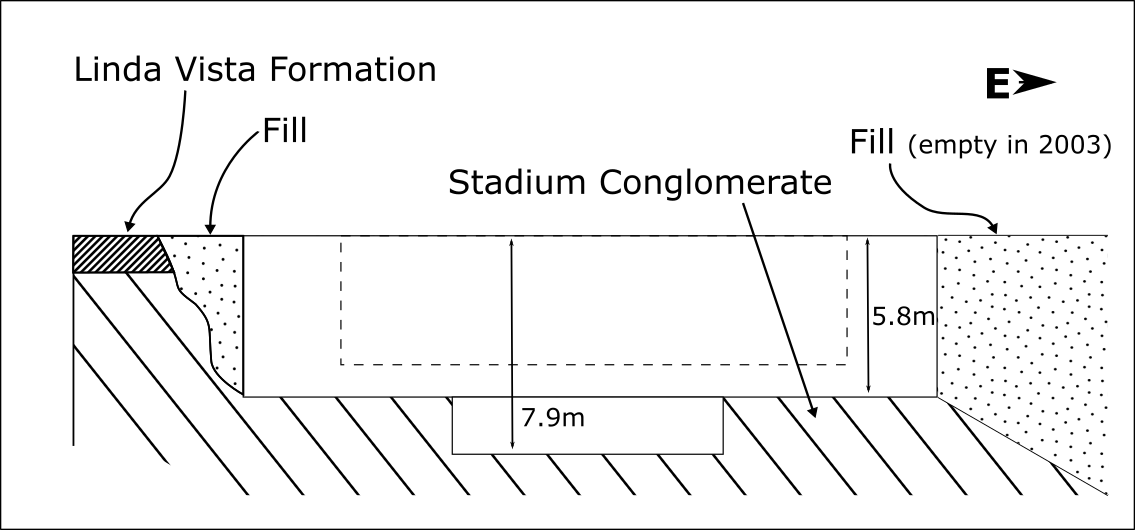} }
 \caption{Section cut displaying the soil profile}
  \label{fig1.3}
\end{figure}

Additional shear wave velocity measurements were performed in 2012. A study conducted by NEES at the University of Texas at Austin reported shear wave velocities overall larger than those presented by \cite{Luco:2011}, with a shear wave velocity of 834m/s for the stadium conglomerate. Furthermore, Post-upgrade CPT tests were performed in 2022 on opposite corners of the reaction mass. Test labeled as CPT-1 was carried approximately 7m away from the northeast corner of the reaction mass, while CPT-2 was performed 2m away from the southwest corner. Soil type results were in accordance with the soil characteristics obtained in previous years. The nominal location of the CPT along with the soil studies previously mentioned are displayed in Figure 1.4. The northeast corner presented clay, silty clay, and silty sand between 0.5m and 6m deep. Very dense/stiff soil corresponding to the stadium conglomerate was found beyond 6m, reaching a maximum depth of 9.36m. The southwest corner presented mostly clay and silty clay between 0.3m and 3m, with the addition of silty sand below 3m indicating the presence of stadium conglomerate at this depth. Very dense/stiff soil was found at between 4.5m and 10.3m and tip refusal was met at 11.3m. The cone resistance $q_c$ found at the bottom of the reaction mass (5.8m) was 5.75MPa (60tsf) and 11.5 (120tsf) for the northeast and southwest respectively. The soil type was overall similar on opposite corners, and the difference at which the dense/stiff soil was found on each corner is congruent with the different fill on each side of the reaction mass, as shown in Figure 1.3.

The CPT results were further used to determine the shear wave velocity based on empirical correlations. The equations were selected following recommendations from Wair, Shantz, and DeJong (\citeyear{Wair:2012}), who proposed using the average of the following correlations by Mayne (\citeyear{Mayne:2006}), Robertson (\citeyear{Robertson:2009}) and Andrus (\citeyear{Andrus:2007}): 

\begin{equation}
    V_s = 118.8 Log(f_s)+18.5 
\end{equation}
\begin{equation}
    V_s=2.62q_t^{0.395}+I_c^{0.912}D^{0.124}SF^a
\end{equation}
\begin{equation}
   V_s=[(10^{(0.55I_c+1.68)})(q_t-\sigma_v)/Pa]^0.5 
\end{equation}

Where $q_t$, $f_s$,$\sigma_v$  ,$P_a$   are the total cone resistance, sleeve friction, vertical stress, and atmospheric pressure, all in kPa, $I_C$ is the Soil Behavior type (SBT) Index, and D is the depth in meters. 

\begin{figure}[H]
 \centering
 {\includegraphics[width=.45\linewidth]{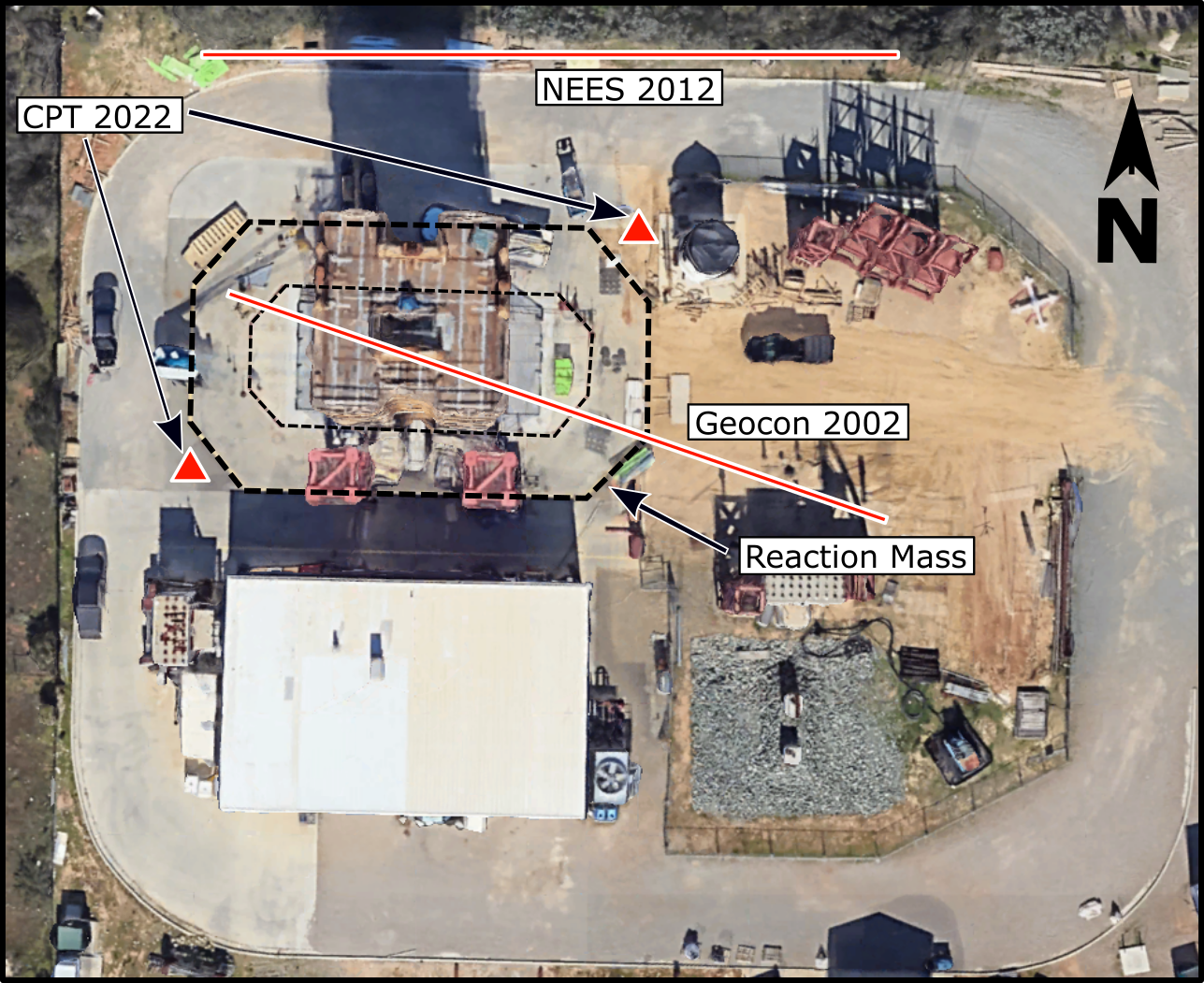} }
 \caption{Satellite view of the facility showing the nominal location of the soil studies carried in since concession of the project in 2002. }
  \label{fig1.4}
\end{figure}

Figure 1.5 shows the shear wave velocity found using the different empirical equations and compares them to the values from 2002 presented by \cite{Luco:2011} and the report from NEES@ UT Austin from 2012. The differences in shear wave velocity displayed in the figure can be explained by the following reasons: the condition of the soil during the tests and location. The study from 2002 was performed on undisturbed naked soil, at the exact location of the reaction mass prior to its construction. Meanwhile, the most recent CPT reflects the results from the infill compacted soil, placed after the construction works were completed. Regarding the location, the measurements reported in 2012 were taken further away from the other studies, approximately 50ft north of the reaction mass on undisturbed naked soil, as seen in Figure 1.4. Results from this study presented larger an overall larger shear wave velocity than the other two. Given the different conditions and locations in which each study was performed, the results cannot be compared among them and are only reported as a range of values to be considered for future computational models. 

\begin{figure}[H]
 \centering
 {\includegraphics[width=.65\linewidth]{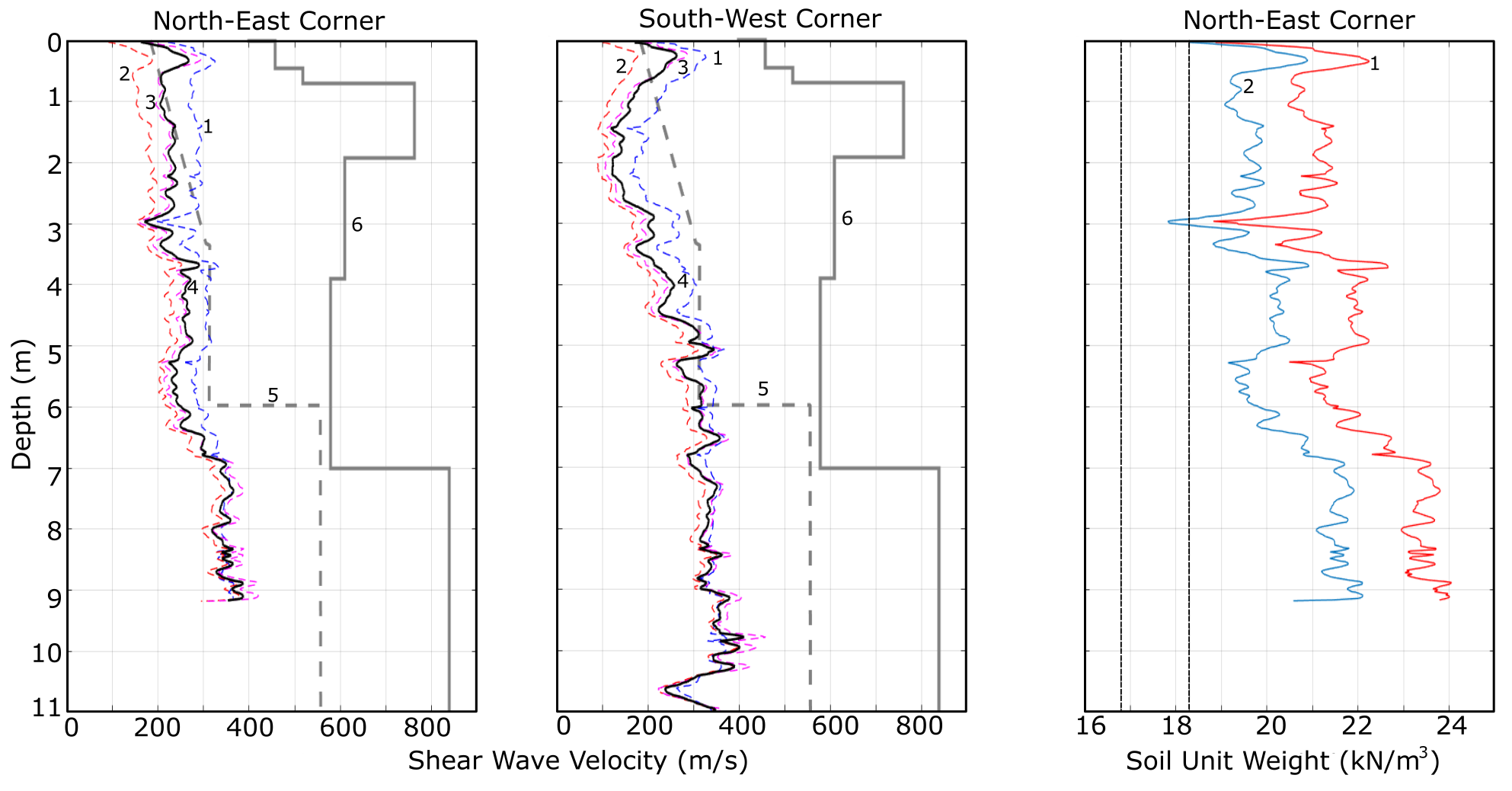} }
 \caption{Shear wave velocity correlated from CPT compared to previous studies. (1) Equation 1.1, (2) Equation 1.2, (3) Equation 1.3, (4) Average, (5) Luco 2011, (6) NEES@ UT Austin 2012. }
  \label{fig1.5}
\end{figure}

\subsection{Forced Vibration Tests Conducted in 2003}

In October 2003, during the construction of phase one, Stepped Frequency tests were carried out to characterize the dynamic response of the reaction mass-soil system. The study took place between October 21-24 during the dry season in San Diego. At the time of the tests, the steel plates and actuators had not been placed yet, and the soil pit to the east of the reaction mass remained empty. The tests were designed to excite the reaction mass in three degrees of freedom independently: East-West (X-DOF), North-South (Y-DOF) and rotation around the vertical axis (Yaw-DOF). 

The forces were applied using the 200 kip capacity NCO MK-a5 eccentric mass shakers owned by UCLA. The shakers were fastened at the top east and west side of the reaction mass. (Nigbor 2012) The shakers had two basket assemblies counter-rotating in a horizontal plane and could be operated synchronously at 0o and 180o phase angles. The eccentric weights placed on the basket produced a unidirectional sinusoidal force whose amplitude was defined as a function of the eccentricity, the weights in the device, and the rotating frequency. \citep{Luco:2011} Consequently, at every excitation with constant frequency, the force amplitude was expected to remain constant as well. It is worth mentioning that the shakers would only emit a signal when there was a shift in frequency but did not record time history data. Thus, the shaker's forcing function was the only source to identify the magnitude of the applied forces. 

Furthermore, the response was recorded with an array of accelerometers distributed in 19 locations around the top and bottom of the structure. Frequency response curves were obtained for the three degrees of freedom tested. These curves show the amplitude of the steady-state response at a given forcing frequency, provided that all results have been scaled to the same forcing amplitude.

Additional forced vibration tests were carried, upon completion of phase one in 2004. In this occasion, dynamic loads were applied by the actuators as the platen was forced to undergo harmonic motions only in the east-west direction. An important aspect of this test was that the force acting on the block was 6.67MN which is 98\% of the nominal maximum actuator force previously described. 

The following dynamic properties were identified from the frequency response curves computed from the tests: translation-plus-rocking modes of vibration in the longitudinal (X-DOF) and transverse (Y-DOF) directions, with a natural frequency of 10 Hz and 10-12Hz for each case, respectively. In addition, the rotational (Yaw) mode of vibration was identified with a natural frequency close to 14Hz. Moreover, the effective damping ratio of the soil-reaction mass system was estimated between 32\% and 42\%. Finally, the high damping of the system and small displacement results validated the unconventional and more economical design of the reaction mass. A thorough analysis of the results, including comparisons with an analytical model, is discussed in \cite{Luco:2011}.

\section{POST-UPGRADE FORCED VIBRATION TESTS}

\subsection{General Description}

A new series of forced vibration tests were performed upon completion of the 6-DOF upgrade. The study was conducted between April 7-20, 2022. This was considered dry season with an observed precipitation of 1.61 inches and 0.65 inches in the months of March and April respectively.

Two types of forced vibration tests, Stepped Frequency and Swept Frequency were applied and are further described in following sections. In addition to the degrees of freedom excited in previous years (X, Y and Yaw), vertical translation (Z-DOF) was also tested using the new set of vertical actuators. During these tests, dynamic loads on the reaction mass were applied by the new configuration of actuators as they moved the platen one DOF at a time in a prescribed harmonic motion. The shake table controller was set to maintain the platen acceleration amplitude constant; however, the controller did not guarantee that the actuator force amplitude would remain constant as well. Moreover, because of the V-shape configuration and the location of the actuators above the center of gravity, the reaction mass was excited in more than one degree of freedom in each test. The effects of these unintended excitations are further discussed in the results section.

\subsection{Testing Setup and Instrumentation}

To record the response of the reaction mass, an array of 31 accelerometers were distributed around the foundation block at three different heights, and two additional sensors were placed directly on the surrounding soil. Figure 2.1 shows the nominal location of the accelerometers measured from the top center of the reaction mass and the coordinate system adopted to present the results.  The three adjacent coordinates indicate that 3-three sensors are located at the same X, and Y coordinates but at different heights. All sensors and brackets were tightly fastened to prevent any undesired vibration. The accelerometers at the top and bottom were directly bolted to the reaction mass, while the ones at mid-height were bolted to a steel bracket and then fastened to the walls of the structure at the desired height, see Figure 2.2. Finally, the coordinate system was chosen in accordance with the tests of 2003 where the east-west direction was aligned with the X-axis.  

The sensors, provided by the Scripps Institute of Oceanography, were manufactured by Kinemetrics. Two types were used: ETNA2 and Episensor, with the second one connected to a Quanterra datalogger to store the data. All sensors' local X and Y-axis were aligned with the adopted global coordinate system. All sensors recorded vibrations at a 200Hz rate continuously during the 2-week duration of the tests, which were then sorted into MATLAB '.m' files \citep{Matlab}. Moreover, applied force time history was recorded by the actuator controller at a 512Hz sample rate. Applied force and response data were then synchronized when analyzing the results.

\begin{figure}[H]
 \centering
 {\includegraphics[width=.75\linewidth]{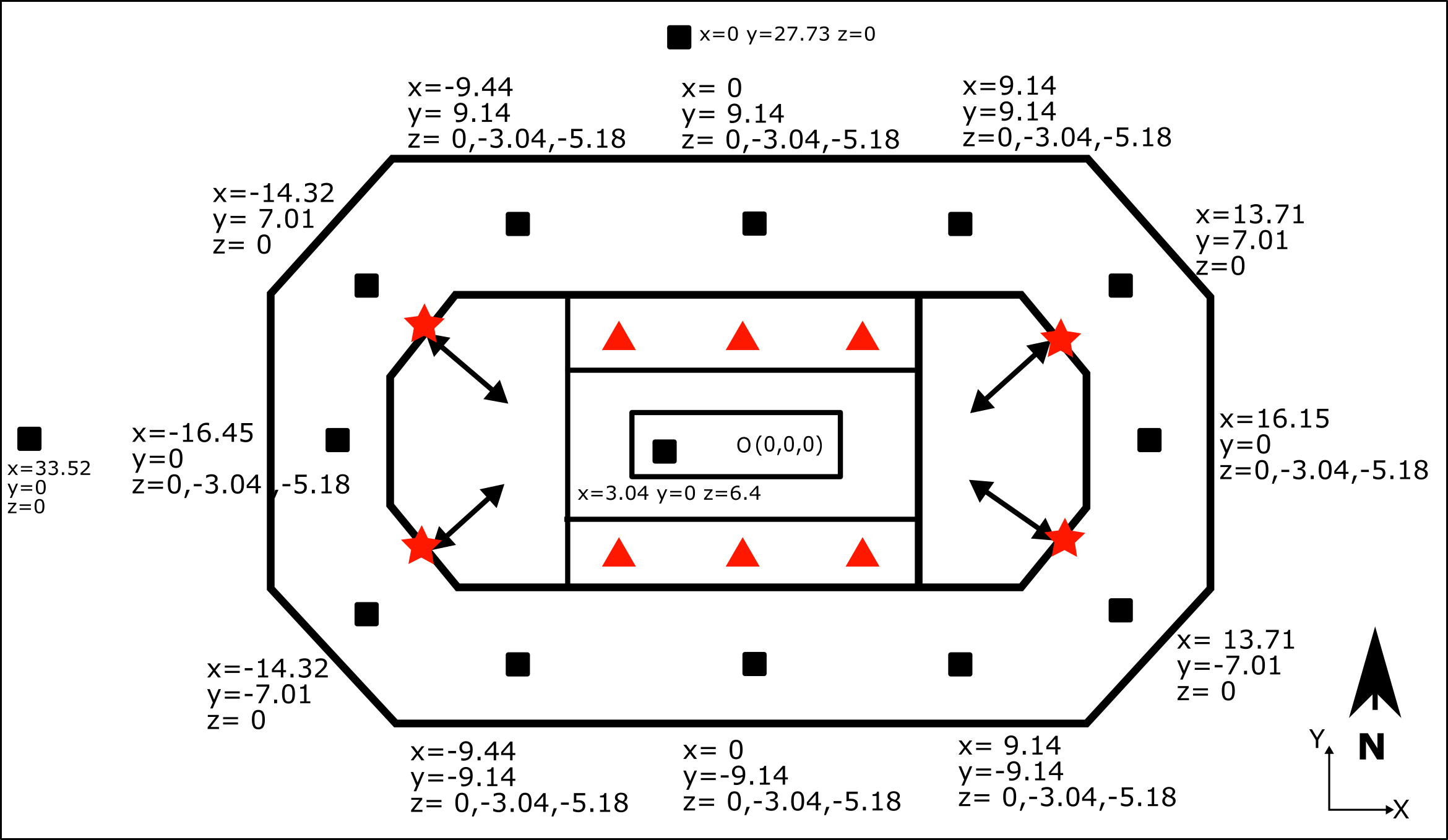} }
 \caption{Test Set-Up showing the location of accelerometers (square), vertical actuators (triangle) and Horizontal actuators (star) which represent the location of the applied forces.}
  \label{fig2.1}
\end{figure}

\begin{figure}[H]
 \centering
 {\includegraphics[width=.65\linewidth]{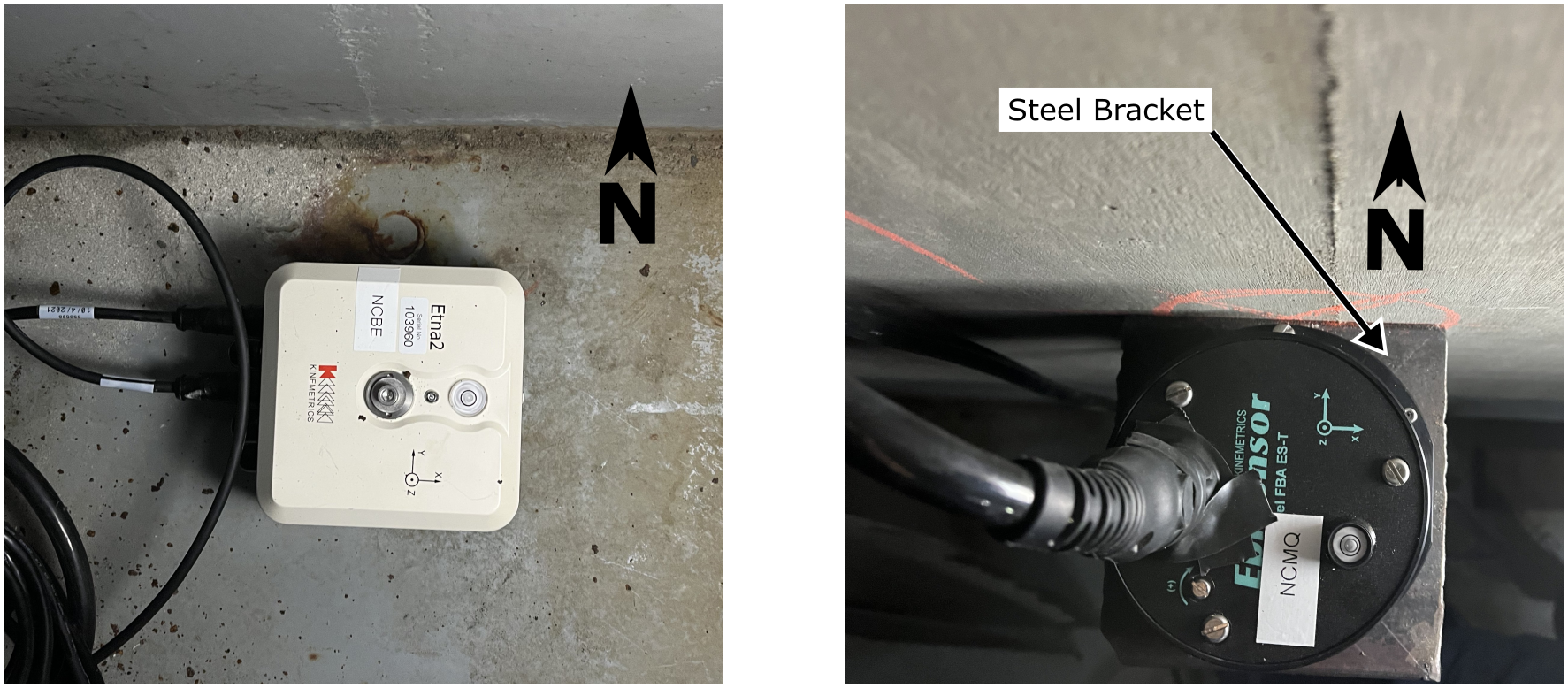} }
 \caption{Fastening of accelerometers at the bottom (left) and mid-height (right) of the reaction mass}
  \label{fig2.2}
\end{figure}

\subsection{Applied Forces: Location and Amplitude}

Given that the controller was commanded to maintain constant platen acceleration and not constant force, the applied forces were carefully inspected before analyzing at the results. The forces applied onto the reaction mass could be determined directly from the actuator controller or from the inertial force caused by the platen acceleration. Nonetheless, it should be noted that the Platen Inertial force was not equal to the total recorded forces, this occurred because the recorded forces not only included the inertial force but also additional forces that the actuators must overcome. These included relatively small internal forces in the mechanism of the actuator and most importantly, forces applied by the pressurized Hold-Down-Struts (HDS) anchored to the bottom of the reaction mass, see Figure 2.3(a). Consequently, to account for all the forces imposed onto the reaction mass, forces recorded by the controller were used as the applied forces.

The HDS forces were further analyzed to determine the magnitude of the reaction these devices could apply, and how this could affect the forced vibration test. Figure 2.3 (a) shows the HDS forces acting as the platen moves, and Figure 2.3 (b) shows the location of the reaction forces applied to the foundation block. During horizontal translation, the HDS imposed a horizontal force that had to be overcome by the actuators. The magnitude of such force depended on the displacement of the platen. The maximum displacement scenario was considered, which occurred during the 1Hz East-West excitation with a displacement of 200mm. At this displacement, each HDS X-component had an approximate magnitude of 100kN. When added together, it represented 22\% of the total recorded force in the X-direction. This meant that during this excitation, a fifth of the force was being applied at the bottom of the reaction mass. At 2Hz, with a maximum displacement of 50mm, the HDS X-component represented only 7\% of the total force and kept decreasing at higher frequencies. Consequently, for frequencies above 2Hz it was considered that the horizontal forces on the reaction mass were applied only at the location of the actuators, and the horizontal component of the HDS could be neglected.

One additional factor was considered when analyzing the applied horizontal forces: the change in the angle between the two east horizontal and two west actuators. As the actuators extended and retracted to move the platen, the angle changed, thus affecting the magnitude of the X and Y components of the actuator force, see Figure 2.3(c). It was found that at the platen displacement of 200mm, the angle was small only had a negligible effect on the computed forces. Therefore, the change in angle was neglected when computing the components of the horizontal actuator forces.

\begin{figure}[H]
 \centering
 {\includegraphics[width=.85\linewidth]{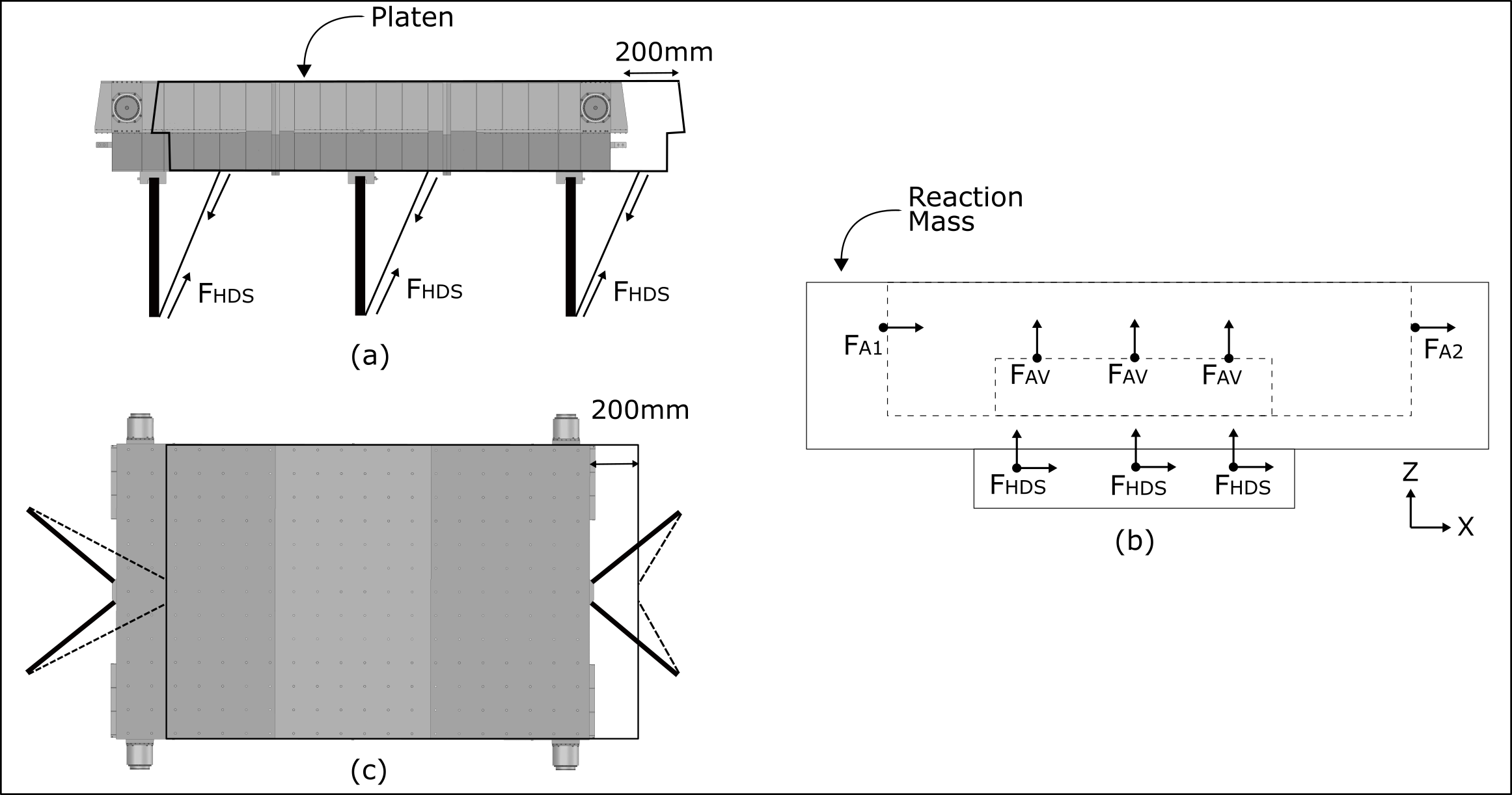} }
 \caption{Additional applied forces: (a) HDS forces occurring at maximum recorded displacement. (b) Location of forces applied onto the reaction mass. FA and FHDS are located at anchoring points of the Actuators and Hold-Down-Struts respectively. (c) Change in V-shape horizontal angle at maximum recorded displacement. }
  \label{fig2.3}
\end{figure}

Regarding the vertical loads, the HDSs were continuously pulling the platen downwards. At the zero position at the beginning of every test, the platen had already raised 0.14m from the rest position. This translation implied a vertical force of over 2600kN per HDS, which had to be overcome by the actuators to move the platen. Consequently, the considerable vertical force applied by the HDS could not be neglected, meaning that vertical forces onto the reaction mass were applied at 9-nine locations: the position of the six vertical actuators and the anchor points of the three hold-down struts at the center-bottom of the reaction mass.   

Furthermore, to gain insight into the magnitude of the applied forces, these were compared to the bearing capacity of the soil. As a first approximation of the applied stress, the total actuator forces were divided by the area of the bottom of the reaction mass, excluding the shear key at the center. In addition, horizontal forces acting above the center of gravity of the structure induced a moment on the soil as well. To evaluate the stress on the soil due to the applied moment, the distribution of stresses was approximated as linear. Considering a soil bearing capacity of 478kPa at a depth of 6m, it was determined that the stress generated by the maximum total force of 1360kN was only 0.45\% of the bearing capacity, meanwhile, the maximum stress induced by a moment was only 0.21\% of the capacity of the soil. 

When looking at the recorded force time history, it was noticed that for some tests, the actuator force amplitude was not kept constant despite the constant acceleration amplitude of the platen. As an example, Figure 2.4 shows the time history of the platen acceleration at 8.5Hz and the corresponding force applied by the north-west actuator. In the figure, it is clearly seen how the shake table controller achieved an almost constant acceleration of the platen while the actuator, on the other hand, did not maintain a constant force, and the time history data clearly shows the presence of a non-intended low-frequency harmonic. Figure 2.5 similarly shows the time history of the platen acceleration and actuator force but this time during the Sweep Frequency tests. In this case, the frequency was increased during the excitation, and it was observed that the controller was not able to perfectly maintain constant platen acceleration throughout the test.

 Nonetheless, when looking at a 5-second window, where the frequency was almost constant with a change of less than 1Hz, the amplitude was almost constant as well. The actuator force amplitude was not constant either, not even during small windows of time, where the existence of low-frequency harmonics could still be observed. The presence of varying force amplitude was more evident when exciting the Yaw and Y degrees of freedom on both types of tests. The implications of non-constant force amplitude and super-harmonics are discussed in the results section.

 \begin{figure}[H]
 \centering
 {\includegraphics[width=1\linewidth]{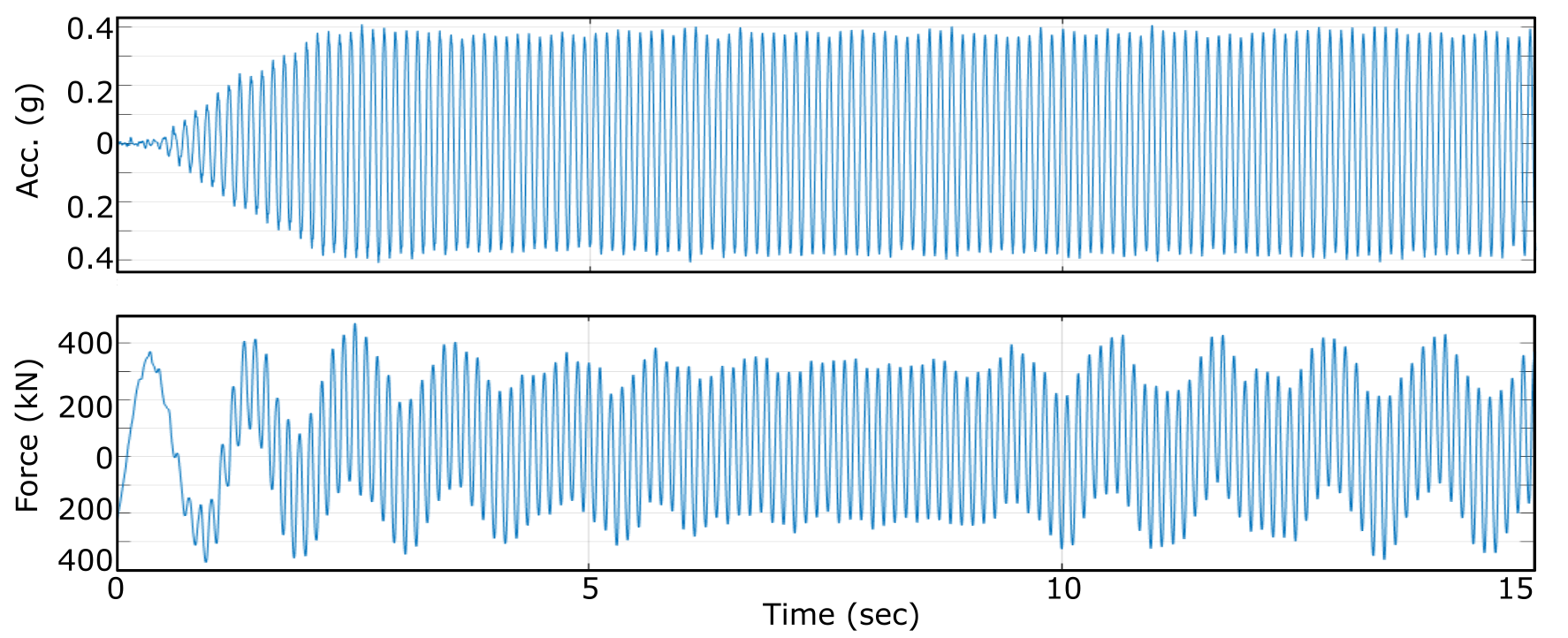} }
 \caption{Time history sample of platen acceleration (top) and Actuator Applied force (bottom) during the Stepped frequency test in the NS direction at 8.5Hz}
  \label{fig2.4}
\end{figure}

\begin{figure}[H]
 \centering
 {\includegraphics[width=.9\linewidth]{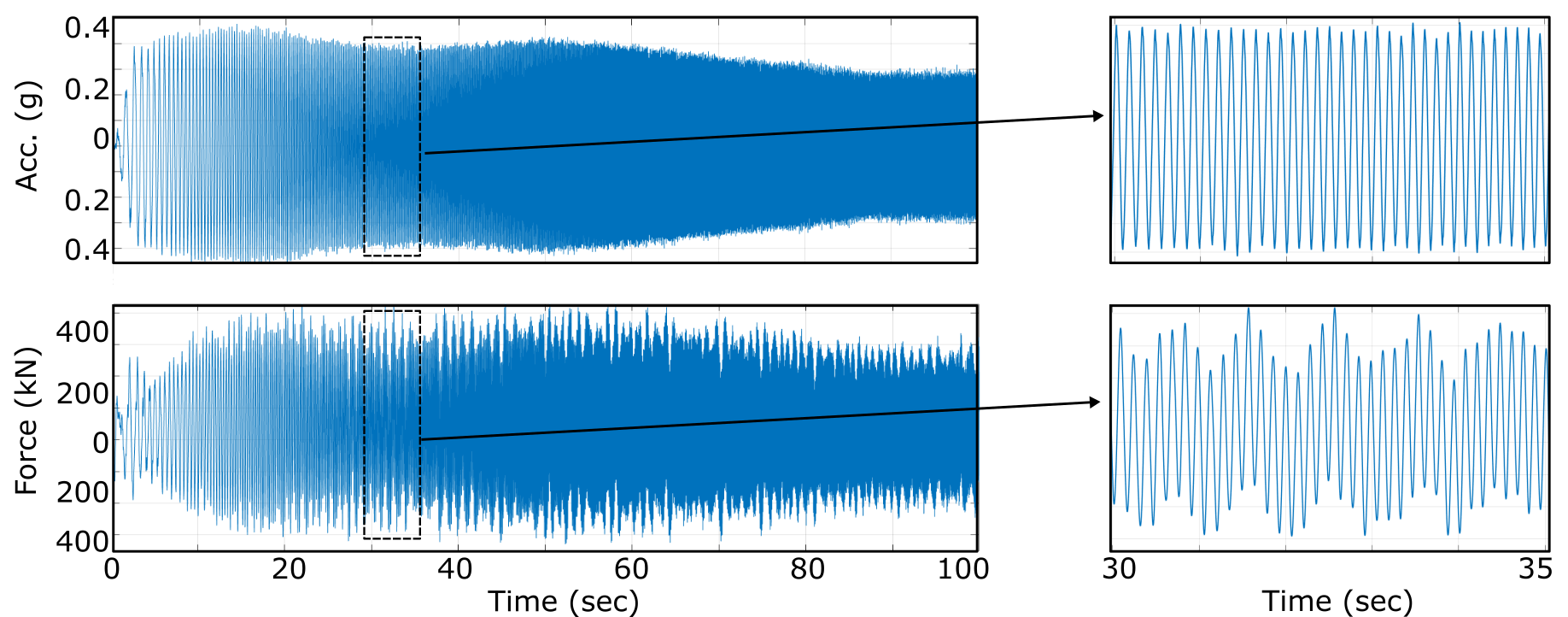} }
 \caption{Time history sample of data from Sweep frequency test in the NS direction: Platen acceleration (top), Recorded actuator Force (bottom) }
  \label{fig2.5}
\end{figure}

\subsection{Stepped Frequency Tests}

In a stepped frequency test, the vibration generator (actuators in this case) is operated at a single chosen frequency and amplitude until the transient response has damped out and the structure's steady-state response can be recorded. Then, the test is stopped, the frequency of the vibration generator is adjusted to a new value, and the procedure is performed once again. This is repeated over a range that covers the frequencies of interest of the system. Subsequently, the steady state displacement amplitude is determined and plotted against its corresponding forcing frequency to obtain the Frequency Response Curve (FRC) of the system. \citep{Chopra:2002}. Finally, the peaks on the frequency response curves are used as an indicator of the natural frequencies of the system 

The stepped frequency procedure was applied to the reaction mass, with the actuators moving the platen one degree of freedom at a time. The frequencies covered a range between 1Hz to 18Hz for the X, Y degrees of freedom, 3Hz to 20Hz for Yaw, and 5Hz to 25Hz for the vertical. Excited frequencies were increased in steps of 1Hz except when close to the expected natural frequency of the system, where the frequency increment was reduced to 0.5Hz.  Table 2.1 summarizes the frequencies covered when performing the tests for each DOF and the minimum and maximum force amplitude recorded across all tests.

The steady-state response of a perfectly linear SDOF system is expected to have constant amplitude and match the frequency of the forcing function, which facilitates the computation of the displacement amplitude and the Frequency response curve. However, the reaction mass-soil system was not perfectly elastic, not even under small vibrations. Thus, finding the acceleration and displacement amplitudes required additional data processing, further described in the results section.

\begin{table}[H]
\caption{Summary of Stepped Frequency Tests.}
 \label{table:assembly1}
\centering
\small
\begin{tabular}{  
>{\centering\arraybackslash}m{5em} 
>{\centering\arraybackslash}m{5em}
>{\centering\arraybackslash}m{16em}
>{\centering\arraybackslash}m{3em}
>{\centering\arraybackslash}m{8.5em} 
} 
\hline\hline
{\textbf{DOF (Direction)}} & {\textbf{Frequency range}} & {\textbf{Frequency increment}} &  {\textbf{\# of Tests}} & {\textbf{Force/torque amplitude range}}  \\
\hline
X (E-W)  & 1Hz-18Hz & 1Hz: between [1-9] Hz and [12-18]Hz  0.5Hz: between [9-12]Hz  & 22 & [1250 – 1360] kN \\

Y (N-S)  & 1Hz-18Hz & 1Hz: between [1-9] Hz and [12-18]Hz  0.5Hz: between [9-12]Hz & 22 & [600 – 714] kN \\

Z (Vert.) & 5Hz-25Hz & 1Hz: between [5-14] Hz and [18-25]Hz  0.5Hz: between [14-18]Hz & 27 & [550 – 780] kN \\

RZ (Yaw) & 1Hz-20Hz & 1Hz: between [1-9] Hz and [12-20]Hz  0.5Hz: between [9-12]Hz & 24 & [2.74–3.16] MN-m \\

\hline
\end{tabular}
\normalsize
\end{table}

\subsection{Sweep Frequency Test}

This procedure is similar to the stepped frequency with one key difference. The test does not stop when changing from one frequency of vibration to the next one. The frequency increase is continuous, and all frequencies in the range of interest are covered in one excitation. Commonly, the increase of the frequency can be linear or logarithmic with the latter being preferred. A linear increase means that all frequencies are excited for the same amount of time before moving to the next. 

Conversely, the logarithmic increase, excites the lower frequencies for a longer time, ensuring that low frequencies are excited enough cycles to reach a steady state. In this case, the linear increase was used at a rate of 0.2Hz per second. Furthermore, it is desired for the force amplitude to remain constant throughout the sweep, but this is not always possible depending on the capabilities of the vibrator. In this experiment, the shake table controller was commanded to maintain the platen acceleration constant, resulting in variations of the actuator force amplitude as the frequency changed, see Figure 2.5. 

The degrees of freedom and frequencies covered were the same as those of the Stepped Frequency. The test was repeated for the same DOF at different platen acceleration amplitudes, thus, allowing us to compare the similar responses at different force amplitudes, and investigate whether results could be scaled linearly based on the applied force. Details of every Swept frequency test as well as the minimum and maximum force amplitude recorded in each test, are listed in Table 2.2  

\begin{table}[H]
\caption{Summary of Sweep Frequency Tests.}
 \label{table:assembly2}
\centering
%\small
%\renewcommand{\arraystretch}{1.25}
\begin{tabular}{
   >{\centering\arraybackslash}m{1em}  
   >{\centering\arraybackslash}m{7em} 
   >{\centering\arraybackslash}m{6em}
   >{\centering\arraybackslash}m{7em}
   >{\centering\arraybackslash}m{7em}
   >{\centering\arraybackslash}m{9em}} 
\hline\hline
{\textbf{\#}} & {\textbf{DOF (Direction)}} &  {\textbf{Frequency Range [Hz]}} & {\textbf{Target Amplitude of Platen accel}} & {\textbf{\% of Peak acceleration Performance}} & {\textbf{Force amplitude range}} \\
\hline
1 & X (E-W)   & 1-18  & 0.19g & 5\%   & [328 – 464] kN\\
2 & X (E-W)   & 1-18  & 0.59g & 15\%  & [915 – 1083] kN\\
3 & X (E-W)   & 1-18  & 0.76g & 20\%  & [1696 – 1880] kN\\
\hline
4 & Y (N-S)   & 1-18  & 0.19g & 5\%   & [368 – 544] kN\\
5 & Y (N-S)   & 1-18  & 0.38g & 10\%  & [852 – 1160] kN\\
\hline
6 & Z (Vert.) & 5-25  & 0.19g & 6\% & [120 – 156] kN\\
7 & Z (Vert.) & 5-25  & 0.6g  & 20\%  & [812 – 1160] kN\\
\hline
8 & RZ (Yaw)  & 1-20  & $9^o/s^2$  & 5\%   & [177 – 490] kN\\
9 & RZ (Yaw)  & 1-20  & $36^o/s^2$ & 20\%  & [390 – 1135] kN\\

\hline
\end{tabular}
\normalsize
\end{table}

\subsection{Simplified Mass-Spring Model of the Reaction Mass}

A simplified rigid mass-spring 3D model was used to simulate synthetic data of a Forced vibration test, which was used to calibrate and evaluate the data processing procedures later used in the experimental results. These included the computation of displacement from acceleration and the Frequency Response Curve (FRC) estimation method. The model yielded direct displacement results that could be compared to those obtained via-acceleration and provided confidence in the numerical procedure applied. Moreover, the model allowed us to isolate variables that could become sources of discrepancy when analyzing the experimental results. Some variables investigated include the presence of noise in the data, signal processing filters, varying amplitude, location of applied force, scaling of results, and the unintended excitation of more than one DOF at a time.

The model was adjusted so that its modal properties and response were within the same order of magnitude as the results from 2003. However, it did not intend to accurately represent the real reaction mass-soil system and was not used to compare experimental vs. computational results. Figure 2.6 shows a 3D view of the mass-spring model. The initial distribution of springs and dashpots, with their respective coefficients, were determined based on current documents used in industry for soil-structure interaction \citep{NIST:2012,Mylonakis:2006}. Although, these documents did not provide guidelines on how to distribute the springs along skewed edges. To maintain consistency with the springs in the X, Y, and Z direction, the springs at the corners were assigned normal to the face of the structure, see Figure 2.6 (b). The backfill soil shown in Figure 1.3 was not considered in the model. Thus, horizontal springs were only distributed around the bottom perimeter of the block and vertical springs were distributed all over the bottom face, see Figure 2.6 (c). To account for rotational stiffness, vertical spring coefficients were amplified at locations closer to the edges. This practice follows the Soil-Structure Interaction guide published by NIST (2012). Finally, the stiffness coefficients were further adjusted ad-hoc to approximate the dynamic properties found from the 2003 tests with no additional iterations, and the damping coefficients were assigned to provide a 37\% damping ratio for all dashpots. 

\begin{figure}[H]
 \centering
 {\includegraphics[width=.8\linewidth]{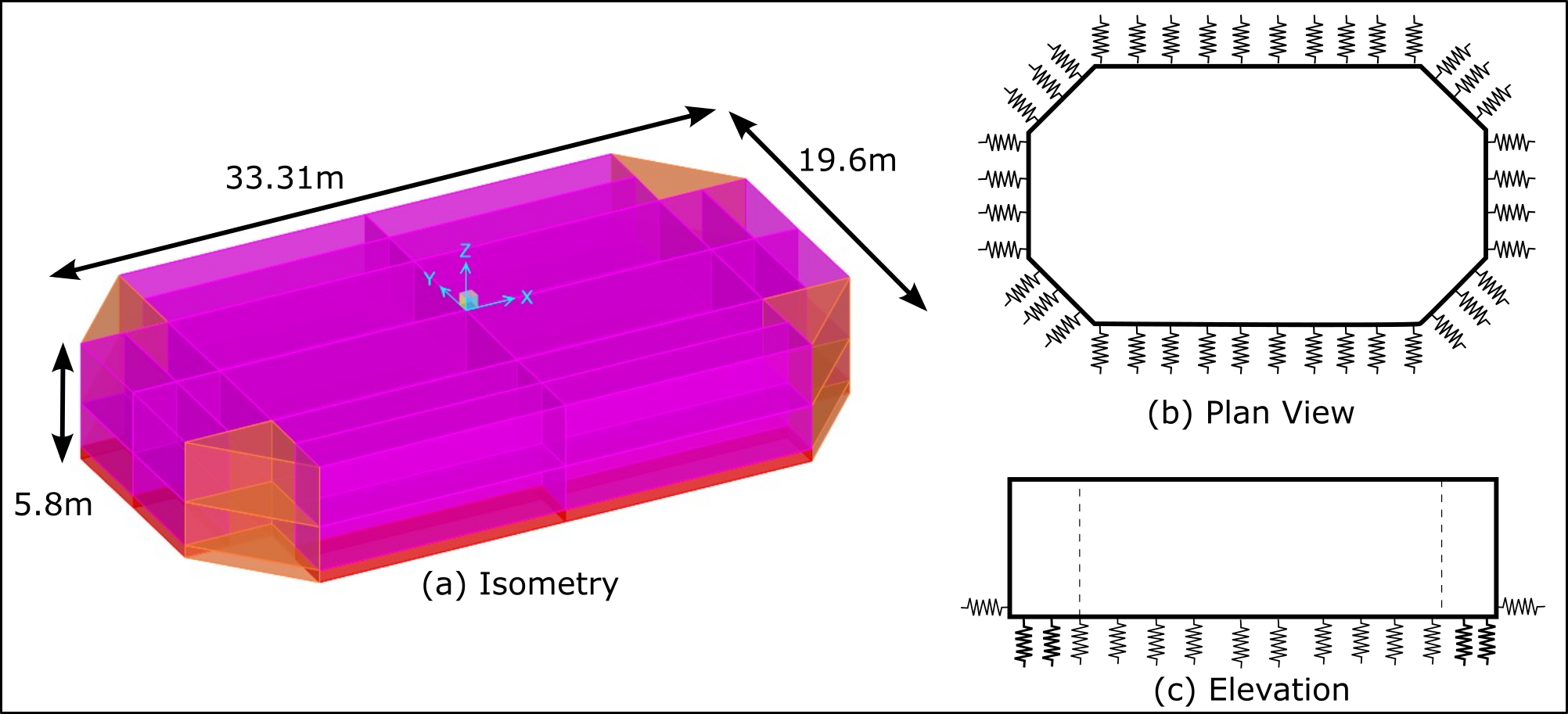} }
 \caption{General view of simplified rigid model. Dimensions of rigid block (left), distribution of vertical and horizontal spring springs at the bottom (right)}
 \label{fig2.6}
\end{figure}

\section{TEST RESULTS}

\subsection{Data Processing}

\subsubsection{Description of Raw Data}

The recorded accelerations were expected to have a similar harmonic response as that of the applied force. In the case of the stepped frequency tests, raw accelerations presented a single frequency response at an almost constant amplitude for all degrees of freedom tested. 

In addition, super-harmonics were expected in the recorded data due to the various sources of nonlinearity in the shake table system, which include the dynamic nonlinearity of the structure-soil system as well as the mechanisms within the servo-valves and the actuators. To provide a deeper explanation: when the shake table operates, the controller transforms the command signal into the achieved table motion. During this process the controller uses the table motion as feedback to compute the commands for the servo-valves and the actuators. The feedback signal contains distortion due to the linear and nonlinear dynamics of the system which cannot be entirely removed by the controller, affecting the output table motion in the form of super-harmonics.

 In addition to the super-harmonics, varying amplitudes were identified, especially in the data recorded during Yaw-DOF excitation.  Figure 3.1 presents a sample of raw acceleration in time and frequency domain, obtained from each of the DOF excited at 7Hz. The response corresponds to a single accelerometer located at mid-height on the north face of the reaction mass. X-DOF response from the East-West excitation, Figure 3.1(a), clearly shows a single harmonic in accordance with the excitation frequency of 7Hz. The response from the North-South and vertical excitation, Figures 3.1(b) and (d), still show a sinusoidal response but the is a clear presence of super-harmonics and more variation in the acceleration amplitude. Nonetheless, the main frequency still matched the frequency of excitation. Finally, data from the Yaw excitation, Figure 3.1(c), showed super-harmonics whose amplitude was greater than the exciting frequency as observed in the frequency domain response. This resulted in acceleration data without a clear main frequency and amplitude. This type of recorded data from the Yaw excitation was only observed in the accelerometers along the centerline of the reaction mass. Stations located towards the edges and corners of the structure recorded data similar to that presented for the other degrees of freedom. 
 
The presence of noise and super harmonics, along with the non-constant force amplitude described in the previous section, influenced the processing of the data described up next. 

\subsubsection{Filtering of Raw Acceleration}

To eliminate the super-harmonics of frequencies not in the range of interest, a band pass filter between 1Hz-25Hz was applied to the raw data. Super-harmonics that occurred when testing low frequencies, such as 14Hz, 21Hz present during the 7Hz test, were not filtered out.  Figure 3.2 shows the frequency response of a fifth degree IIR Butterworth filter with corner frequencies at 1Hz and 25Hz. Despite using an IIR filter with non-linear phase, when given the entire dataset, MATLAB applies a zero-phase filtering approach to eliminate the nonlinear phase distortion. The phase v. frequency plot in Figure 3.2 shows a close to linear phase in the range of frequencies of interest. The nth degree bandpass filter requires 2n+1 coefficients and is defined by the transfer function:

\begin{equation}
    H(z)=\frac{B(z)}{A(z)} = \frac{b_1+b_2z^{-1}+...+b_{n+1}z^{-n}}{a_1+a_2z^{-1}+...+a_{n+1}z^{-n}} 
\end{equation}

\begin{figure}[H]
 \centering
 {\includegraphics[width=.95\linewidth]{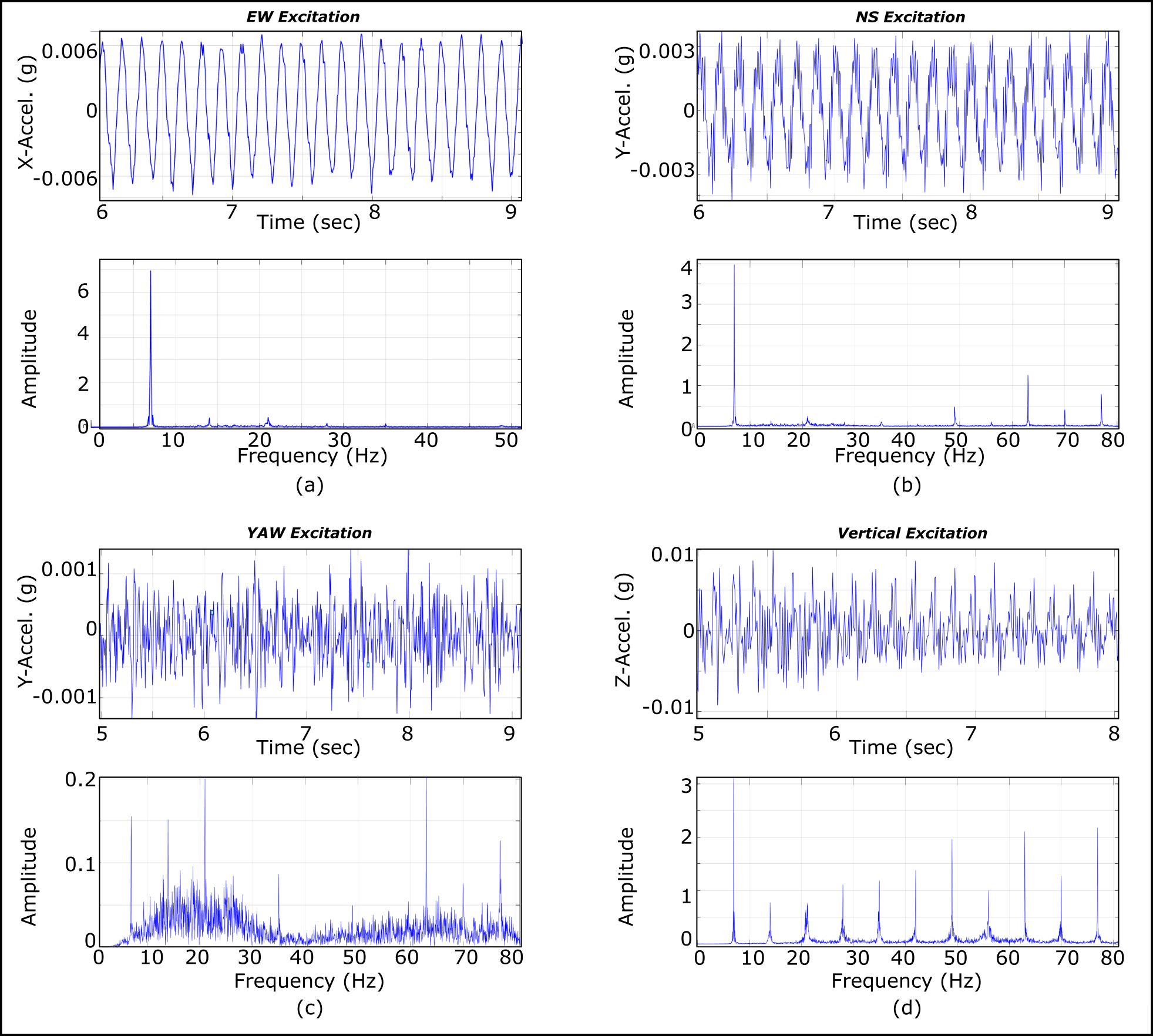} }
 \caption{Sample of unprocessed acceleration data in the time and frequency domain recorded by a single station located on the north face at mid-height. Response to: (a) East-West, (b) North-South, (c) Yaw, and (d) Vertical excitation.  }
  \label{fig3.1}
\end{figure}

\begin{figure}[H]
 \centering
 {\includegraphics[width=.7\linewidth]{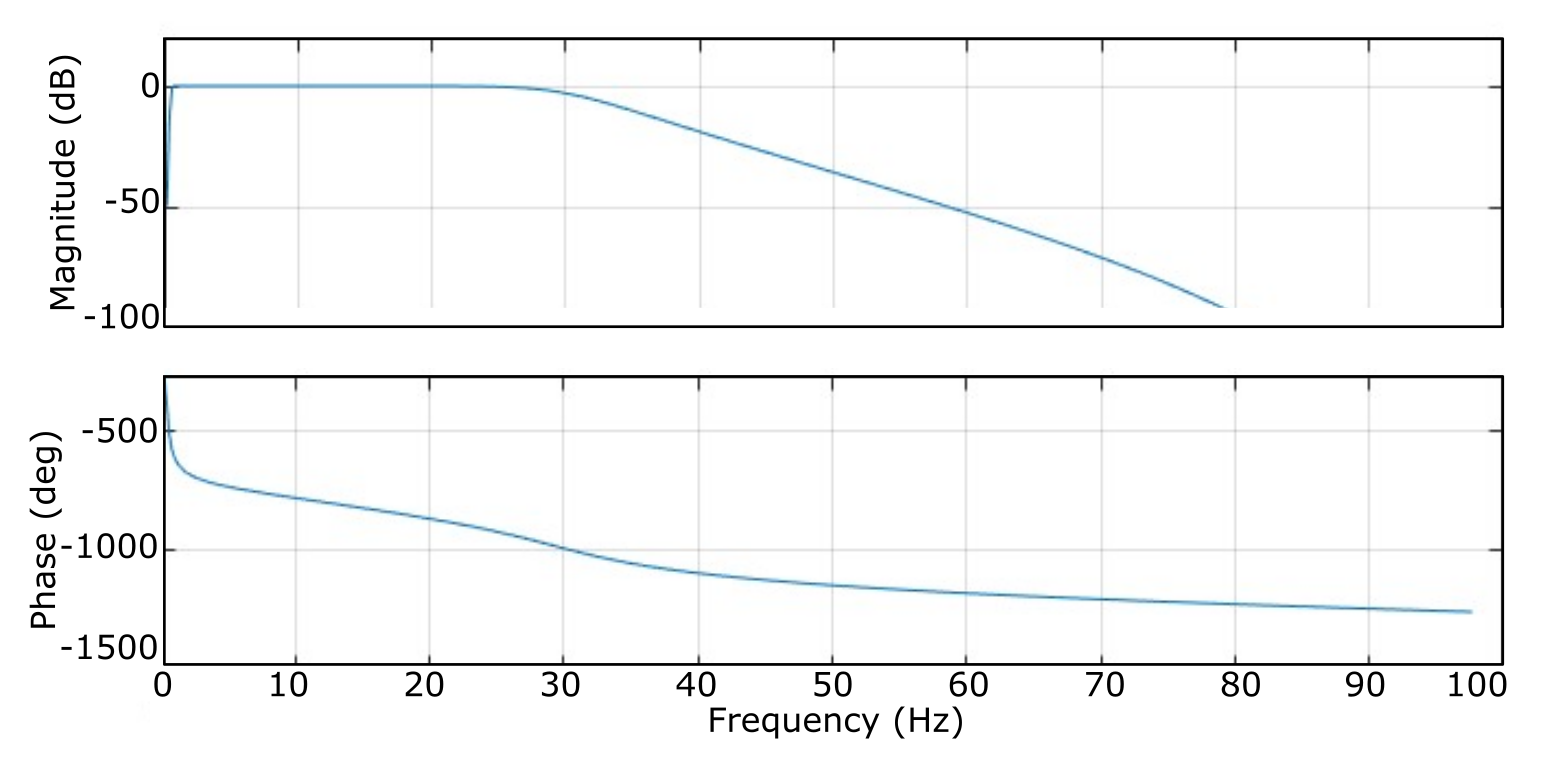} }
 \caption{Bandpass filter frequency response: Magnitude (top) and phase (bottom)}
  \label{fig3.2}
\end{figure}

\subsubsection{Sinusoidal Least Square Fit (LSF)}

Least Square Fit is a technique used to find the coefficients of a function that minimize the square of the difference between the set of measured and estimated data. The difference to be minimized is defined by the objective function:

\begin{equation}
    J= \sum_{i=1}^{N}(\ddot{u}_{i\_measured}-\ddot{u}_{i\_LSF})^2 
\end{equation}

Where $\ddot{u}_{i\_measured}$, $\ddot{u}_{i\_LSF}$ are the measured and estimated acceleration values at time $i$ respectively. In this case the data of interest was the steady state acceleration response, expected to be a single frequency harmonic. Thus, a sine function $\ddot{u}_{LSF}=A1 sin(A_2 t + A_3)$ was selected to approximate the results, where the A coefficients correspond to the amplitude, angular frequency, $\omega$, and phase, $\phi$, respectively. A portion of the steady state response of the recorded data was used as the input. For the steeped frequency results, this window of data could reach a length of 40 seconds. The sweep frequency data on the other hand, could only reach a window of about 2 seconds since the forcing frequency was constantly increasing. With this technique it was possible to obtain acceleration data with infinite resolution which was easily integrated into displacement $u_{LSF}=-A_1 / \omega^2 sin(\omega+\phi)$.

In most cases the recorded response was a clear single frequency harmonic, which was easily fitted with the sinusoidal LSF, see Figure 3.3 (left). However, data with the presence of super-harmonics could not be perfectly fitted, yet the main frequency could be identified with the amplitude that minimized the error, as shown in Figure 3.3 (right) 

The procedure was applied on all accelerometer records in all directions and all forcing frequencies. Finally, the displacement amplitude values  $A_1/\omega^2$ at every forcing frequency were used to obtain the frequency response curves. 

\begin{figure}[H]
 \centering
 {\includegraphics[width=.8\linewidth]{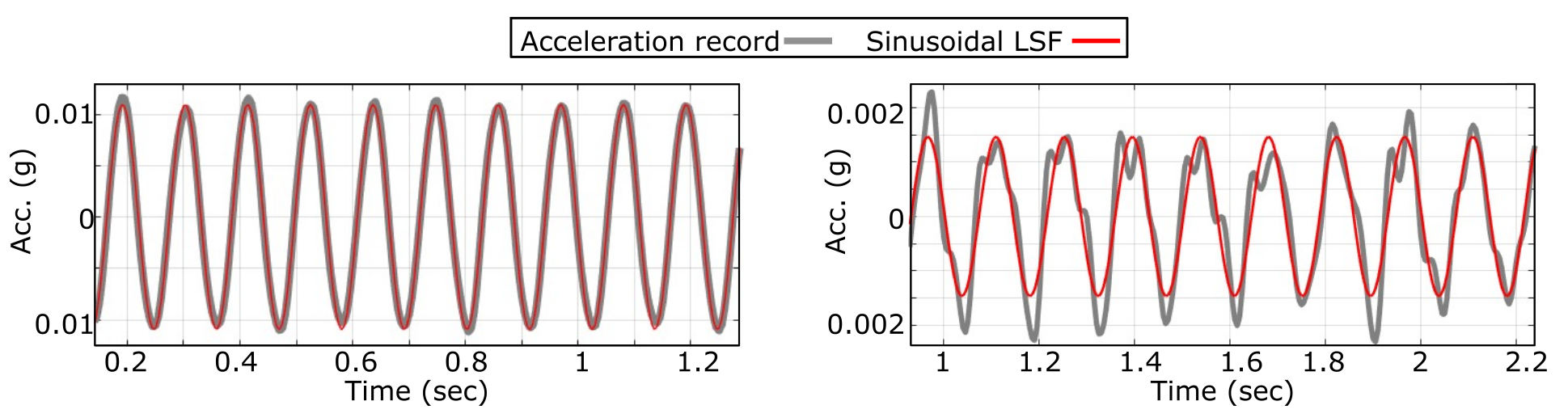} }
 \caption{Sample of steady-state acceleration record (grey) and function obtained via LSF (red): Response to EW excitation at 10Hz (left), response to Yaw excitation at 7Hz (right)}
  \label{fig3.3}
\end{figure}

\subsubsection{Computing Total Displacements}

From the LSF procedure previously explained, the displacement amplitude values  $A_1 / \omega^2$ at every forcing frequency were used to obtain the frequency response curves. It should be emphasized that experimental results correspond to the total displacement composed by the Rigid Body Motion of the block-soil system and the deformational component. These components of the total displacement were later analyzed separately.

The numerical procedure was checked with the synthetic data from the simplified model and its frequency response curves are shown in Figure 3.4, where they are compared to the true displacement values of the model. The figures present a good agreement between the displacement amplitude obtained via LSF and the benchmark data for both types of tests, providing confidence in the methods applied for data processing. Nonetheless, it should be noted that in the case of the stepped frequency test, for low frequencies, below 6Hz, the agreement between the benchmark and the LSF amplitude is not as good. This observation was considered when analyzing the experimental results presented in the following sections.   

\begin{figure}[H]
 \centering
 {\includegraphics[width=.8\linewidth]{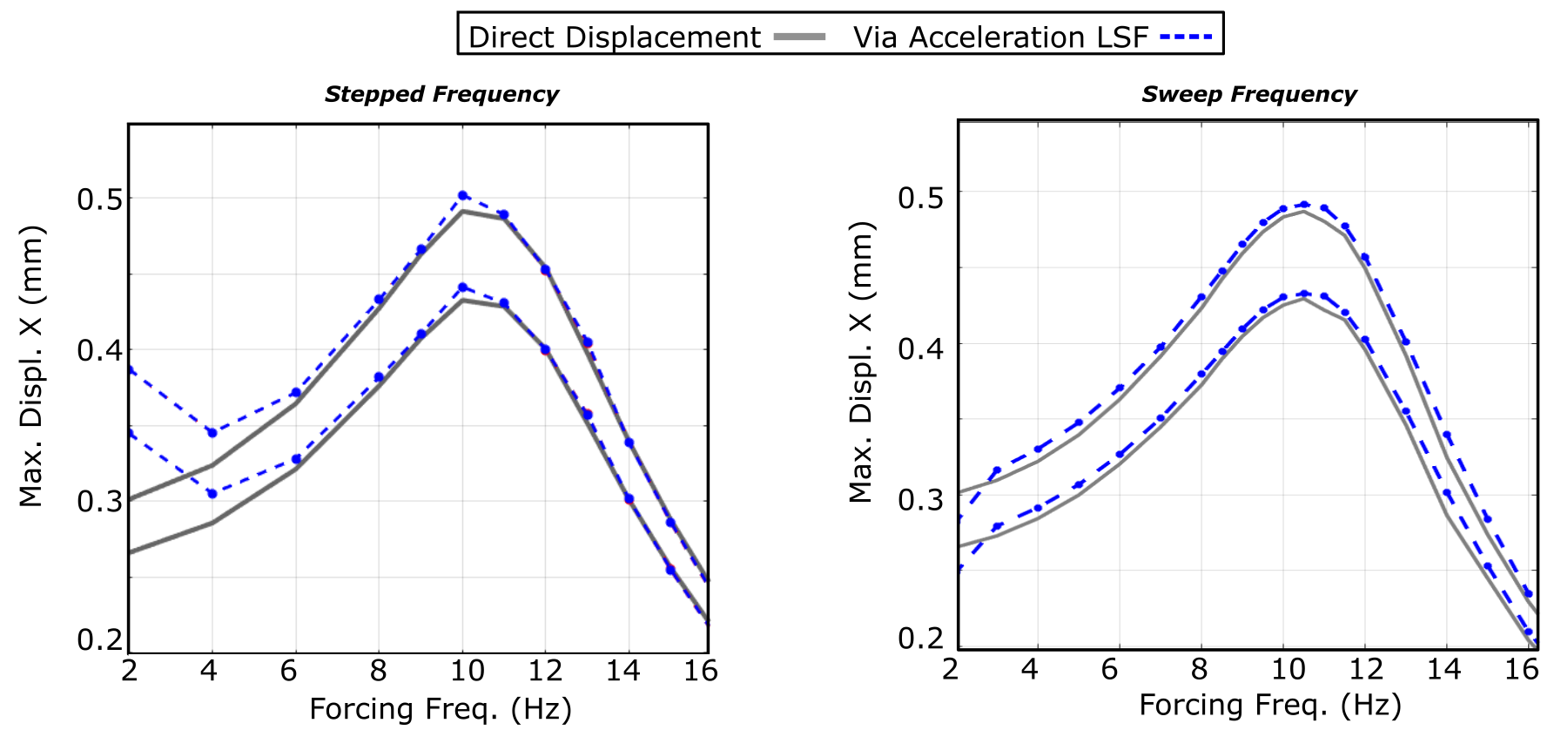} }
 \caption{Frequency Response Plot of simplified rigid model subjected to Stepped-Frequency Tests (left) and sweep frequency tests (right). Displacement output from model (Grey), Displacement obtained from Sinusoidal LSF (Blue)}
  \label{fig3.4}
\end{figure}

\subsubsection{Computing Rigid Body Displacement}

To further analyze the displacement results, contribution of the Rigid Body Motion (RBM) was investigated. The average RBM corresponds to the displacement that the sensors would experience if the structure moved as a rigid block. In rigid body motion, all displacements can be found based the displacement $\Delta_0={\Delta_x,\Delta_y,\Delta_z,\theta_x,\theta_y,\theta_z}$ of a reference point $P_0$ and the location of the sensors with respect to this point.

This way, the rigid body displacement at $i^{th}$ station  $v_i$ is defined by: 

\begin{equation}
    v_i= \begin{Bmatrix} v_{ix} \\ v_{iy} \\ v_{iz} \end{Bmatrix}
    =  \begin{bmatrix} R \end{bmatrix} \begin{Bmatrix} x_i \\ y_i \\ z_i  \end{Bmatrix}
    + \begin{Bmatrix} \Delta_i \\ \Delta_i \\ \Delta_i  \end{Bmatrix} -
      \begin{Bmatrix} x_i \\ y_i \\ z_i  \end{Bmatrix}
\end{equation}

\begin{equation}
R = \begin{bmatrix}
    C\theta_y \ C\theta_z & C\theta_z\ S\theta_x\ S\theta_y - C\theta_x\ S\theta_z\  &
    C\theta_x\ C\theta_z\ S\theta_y + S\theta_x\ S\theta_z\ \\
    
    C\theta_y \ S\theta_z & C\theta_x\ C\theta_z\ + S\theta_x\ S\theta_y\ S\theta_z\  &
    -C\theta_y\ S\theta_x + C\theta_y\ S\theta_y\ S\theta_z\ \\

    -S\theta_y            & C\theta_y\ S\theta_x\  & C\theta_x\ C\theta_y
    \end{bmatrix}
\end{equation}

Where $R$ is the 3D rotation matrix, $C\theta=cos(\theta)$ , and $x_i,y_i,z_i$ are the coordinates of the station with respect to $P_0$.

Given that the expected rotations were small, the rotation matrix was linearized, and the expression reduced to equation (3.5) for a single station and (3.6) for all stations at once:

\begin{equation}
    v_i= \begin{bmatrix}
         1 & 0 & 0 & 0    & z_i  & -y_i\\
         0 & 1 & 0 & -z_i & 0    & x_i\\
         0 & 0 & 1 & y_i  & -x_i & 0
         \end{bmatrix} 
         \begin{Bmatrix}
         \Delta_x \\ \Delta_y \\ \Delta_z \\ \theta_x \\ \theta_y \\ \theta_z    
         \end{Bmatrix}
         = \begin{bmatrix}\alpha_i\end{bmatrix}\begin{Bmatrix}\Delta_0\end{Bmatrix}
\end{equation}

\begin{equation}
    v= \begin{bmatrix} \alpha_1 \\ \alpha_2 \\ ... \\ \alpha_N \end{bmatrix} 
       \begin{Bmatrix} \Delta_0\end{Bmatrix}
\end{equation}

Where i=1 to N=29 corresponds to the sensors in the reaction mass excluding those on the surrounding soil. 

With the use of a Least square fit, it was possible to find the elements of $\Delta_0$ that minimized the difference between the experimental displacement data $u_{LSF}$, and the estimated displacements $v_i$. The found coefficients correspond to the average rigid body motion of the block. The objective function to be minimized in the LSF was defined as:

\begin{equation}
    J= \sum_{i=1}^{N}(|u_{i\_LSF}-v_i|)^2 
\end{equation}

\subsection{Stepped Frequency Results}

This section presents the frequency response curves and deformation pattern of every degree of freedom tested, as well as the estimated rigid body response. Rigid body motion was estimated based on the results from all stations and provided an insight of the overall response of the system concentrated at a single point. 

Given the number of sensors placed on the reaction mass, accelerometer results were combined into groups whose average displacement was plotted. The accelerometers in each group were selected to maintain resemblance with the configuration from the 2011 study \citep{Luco:2011}, with exception of group B2. Figure 3.5 presents the markers used to distinguish each group and the nominal location of the accelerometers. These names and markers are used throughout the results section of the report.

Moreover, to compare results from rotational and translational degrees of freedom, rotation results were multiplied by $l=16.5m$ which is half the length of the reaction mass. Finally, to compare the displacement among separate tests in the frequency response plots, all displacement results were linearly scaled to the same force amplitude of 6800 kN. As previously explained, this was the amplitude used to scale the results from 2003 since it corresponds to the nominal maximum actuator force.

\begin{figure}[H]
 \centering
 {\includegraphics[width=.8\linewidth]{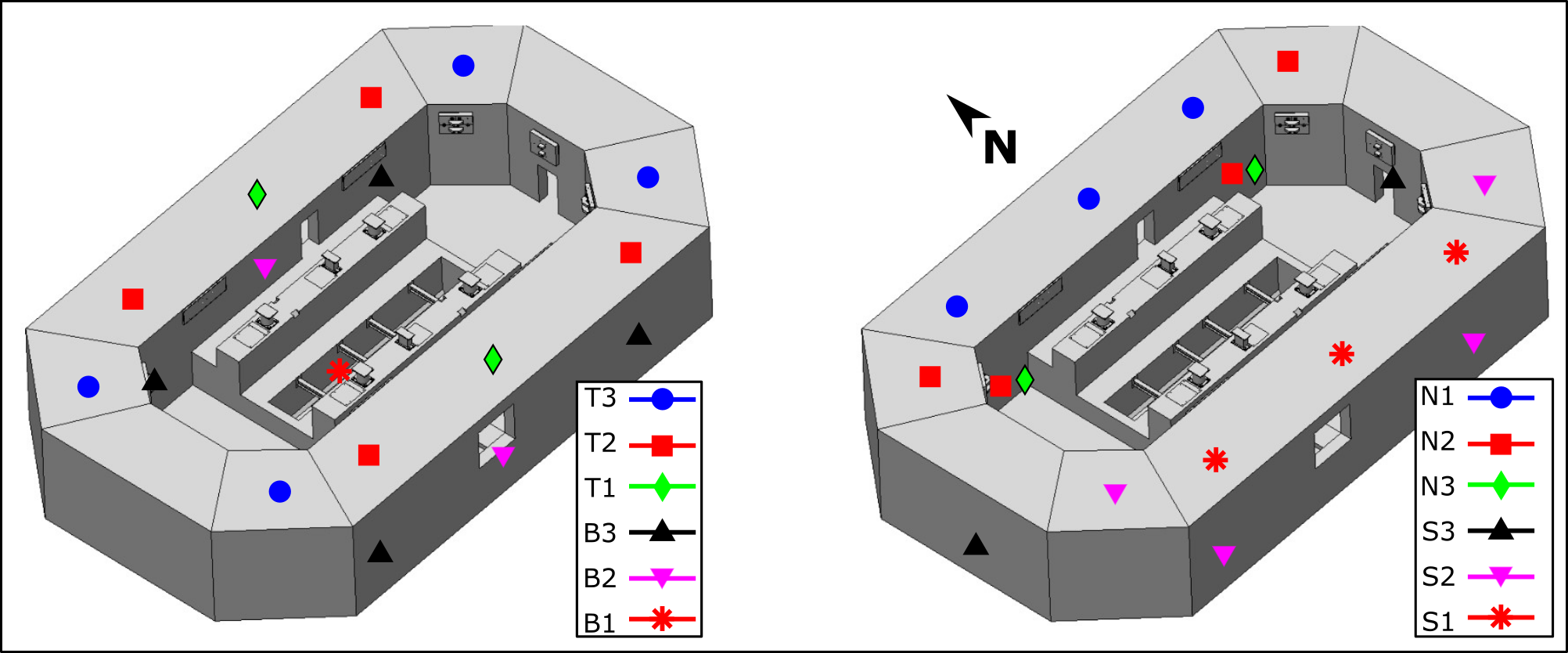} }
 \caption{Accelerometer group names and markers used for the average displacement amplitude presented on the frequency response curves.}
  \label{fig3.5}
\end{figure}

\subsubsection{East-West Excitation (X-DOF)}

Figure 3.6 presents the frequency response curves obtained from exciting the reaction mass in the East-West Direction. In this figure, the 29 stations were grouped based on their location, see Figure 3.5, and only the average of each group was plotted for ease of visualization. Figure 3.6 (a) presents a maximum horizontal displacement amplitude beteen 9.5-10Hz suggesting that this is the fundamental frequency for this degree of freedom. A similar behavior is observed in the vertical displacement, Figure 3.6(b), indicating that the response of the block was a combination of translation and rotation around the Y-axis (pitch). Furthermore, Figure 3.6(c) presents the frequency response curve of the estimated rigid body motion of the reaction mass, which resembles the total displacement, and no localized behavior was identified. In addition, the displacements due to rocking around Y-axis corroborate the presence of pitch in the mode of vibration. To gain deeper insight, rigid body displacement was compared to total displacement on every accelerometer. The average from all stations showed that   94\% and 89\% of the total horizontal and vertical displacement respectively was due to rigid body motion. These percentages correspond to the response at a 10Hz excitation. However, the calculations were repeated with results from all the other frequencies. The results considering all frequencies tested are presented at the end of this section.

\begin{figure}[H]
 \centering
 {\includegraphics[width=.8\linewidth]{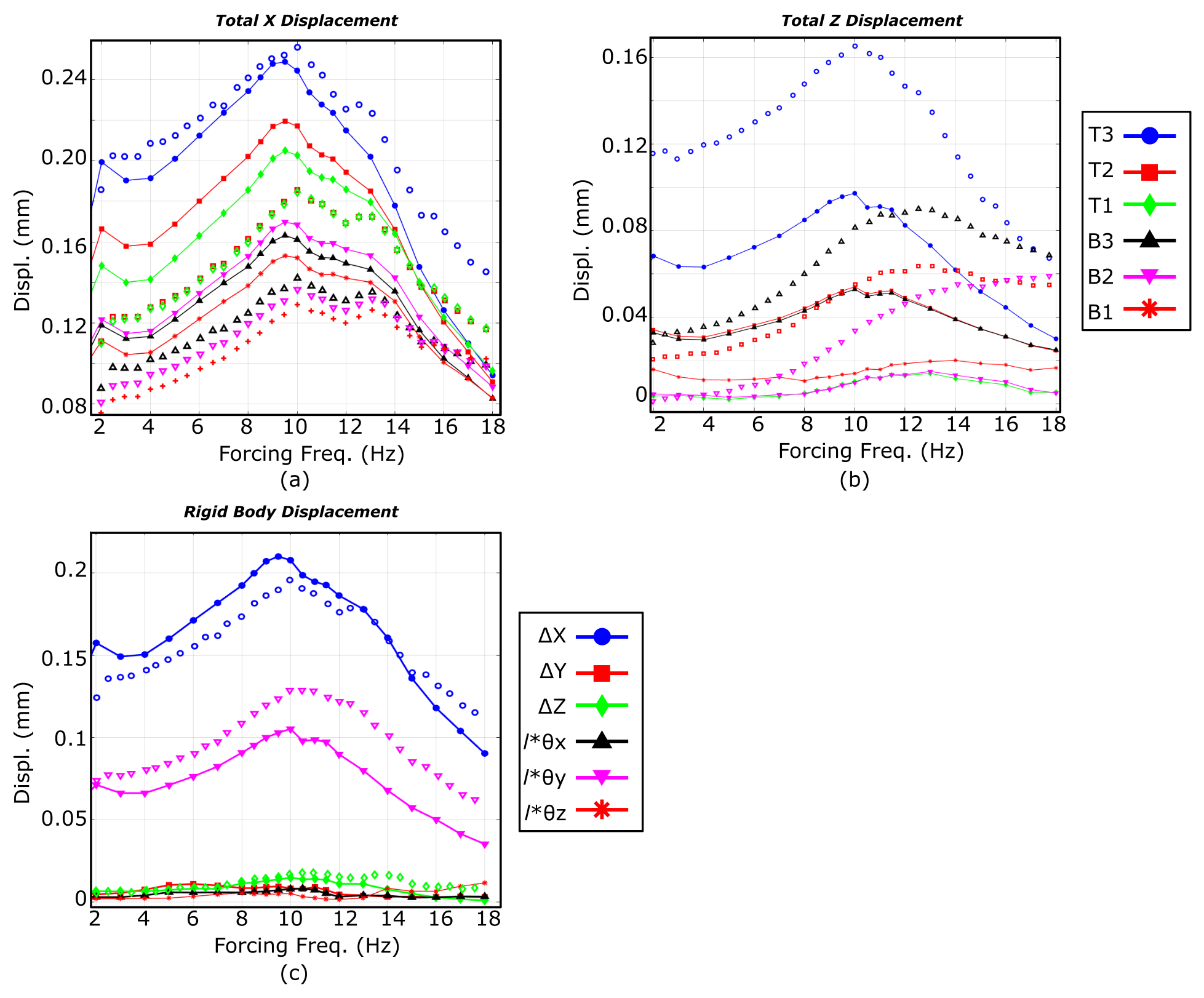} }
 \caption{Frequency Response Curve of East-West Excitation (X-DOF). Comparison of Stepped Frequency tests (solid marker), 2003 tests (open marker) : (a) Horizontal total displacement, (b) Vertical total displacement, (c) Rigid Body Displacement. }
  \label{fig3.6}
\end{figure}

Figures 3.7, 3.8 show the initial location of the stations in the reaction mass and its deformation pattern when excited at 10Hz. The observed deformation patterns were congruent with the frequency response curves, where larger horizontal displacement resulted in large vertical displacement as well. In these figures one can appreciate the contribution of rigid body displacement to the total deformation of the block. It was observed that the top of the reaction mass, Figure 3.7(a), had larger deformation while the response at the middle and bottom of the structure is mostly rigid body translation. Moreover, the deformed shape in Figure 3.7(a) shows the east corner stations moving outward in the Y direction while the west corner moved inward. This deformation is congruent with the direction of the applied forces and corroborates that the reaction mass was excited in other degrees of freedom in addition to the one of interest.

\begin{figure}[H]
 \centering
 {\includegraphics[width=.8\linewidth]{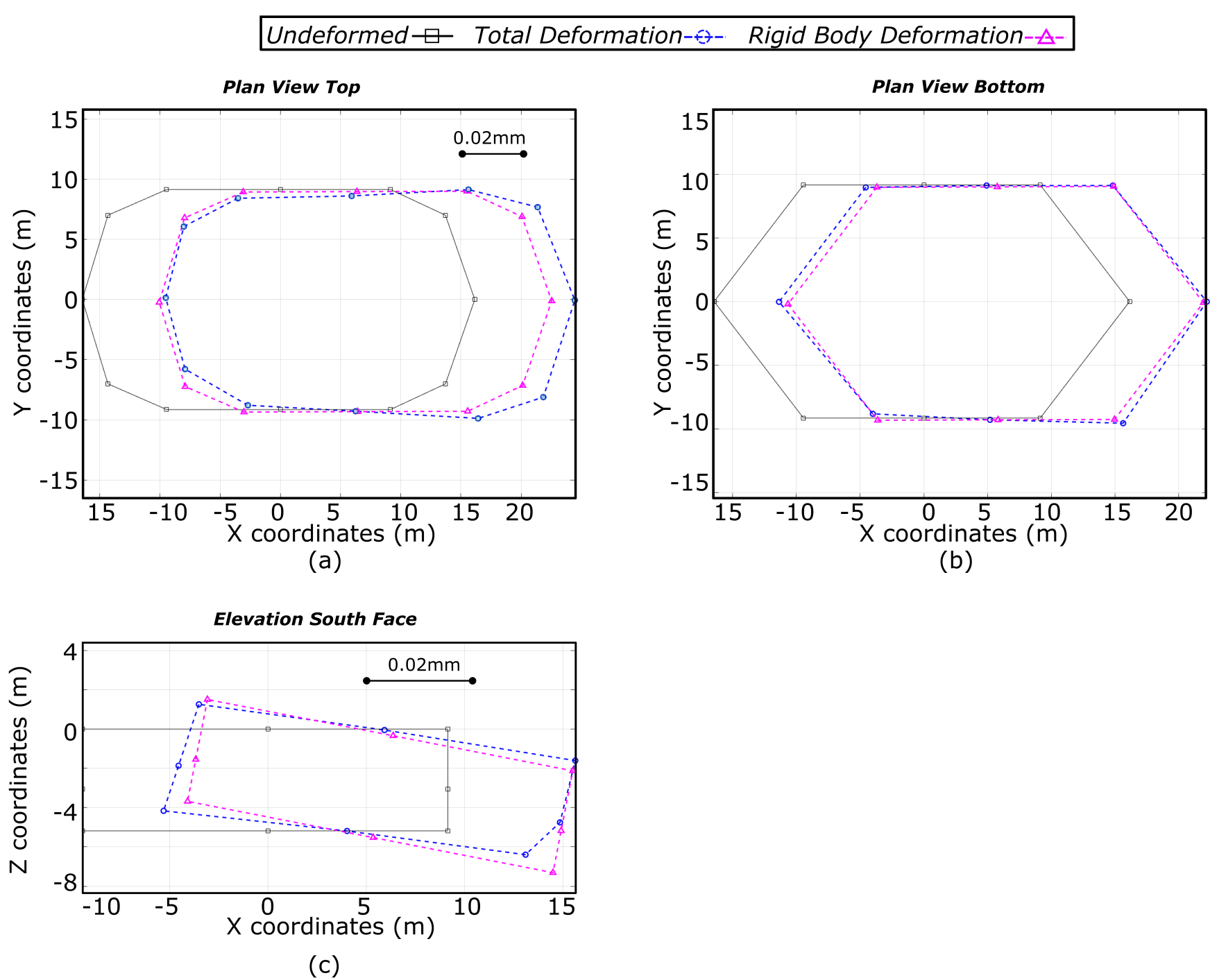} }
 \caption{Deformed configuration showing Total displacement and rigid body displacement from East-West Excitation (X-DOF) at 10Hz: Top of reaction mass (a), bottom of reaction mass (b), South face elevation (c)}
  \label{fig3.7}
\end{figure}

\begin{figure}[H]
 \centering
 {\includegraphics[width=.8\linewidth]{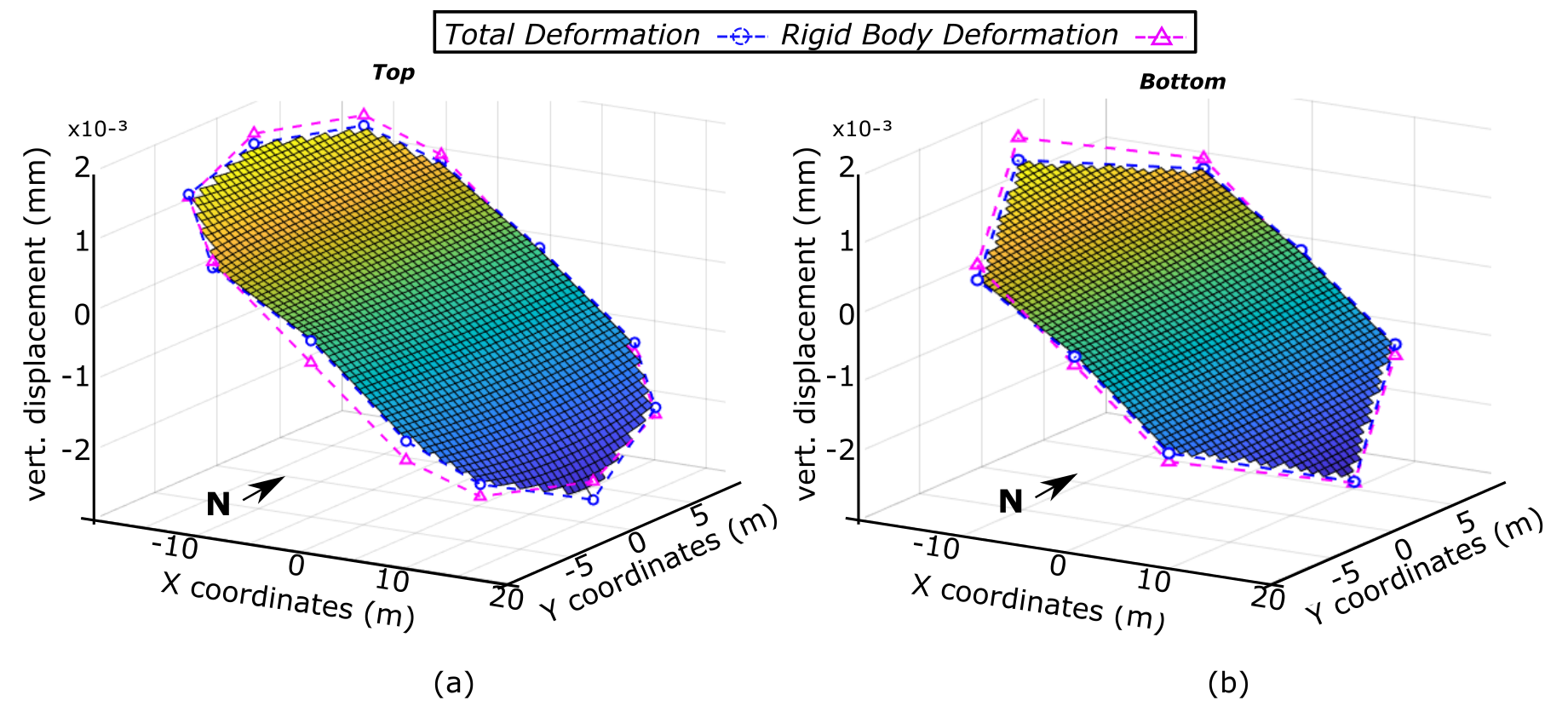} }
 \caption{Total vertical displacement (square) and rigid body vertical displacement (triangle) from East-West Excitation (X-DOF) at 10Hz: Top of reaction mas (a), bottom of reaction mass (b)}
  \label{fig3.8}
\end{figure}

Finally, the deformation pattern was computed in opposite directions to analyze the symmetry of the response. It was observed that the horizontal response was almost the same regardless of the direction. However, the vertical response was larger on the east side of the reaction mass, see Figure 3.9. This can be explained by the different soil profiles at the opposite ends of the reaction mass. The west side was exposed to the original Stadium conglomerate soil while the east side was backfilled with less stiff soil as depicted in Figure 1.3

\begin{figure}[H]
 \centering
 {\includegraphics[width=.8\linewidth]{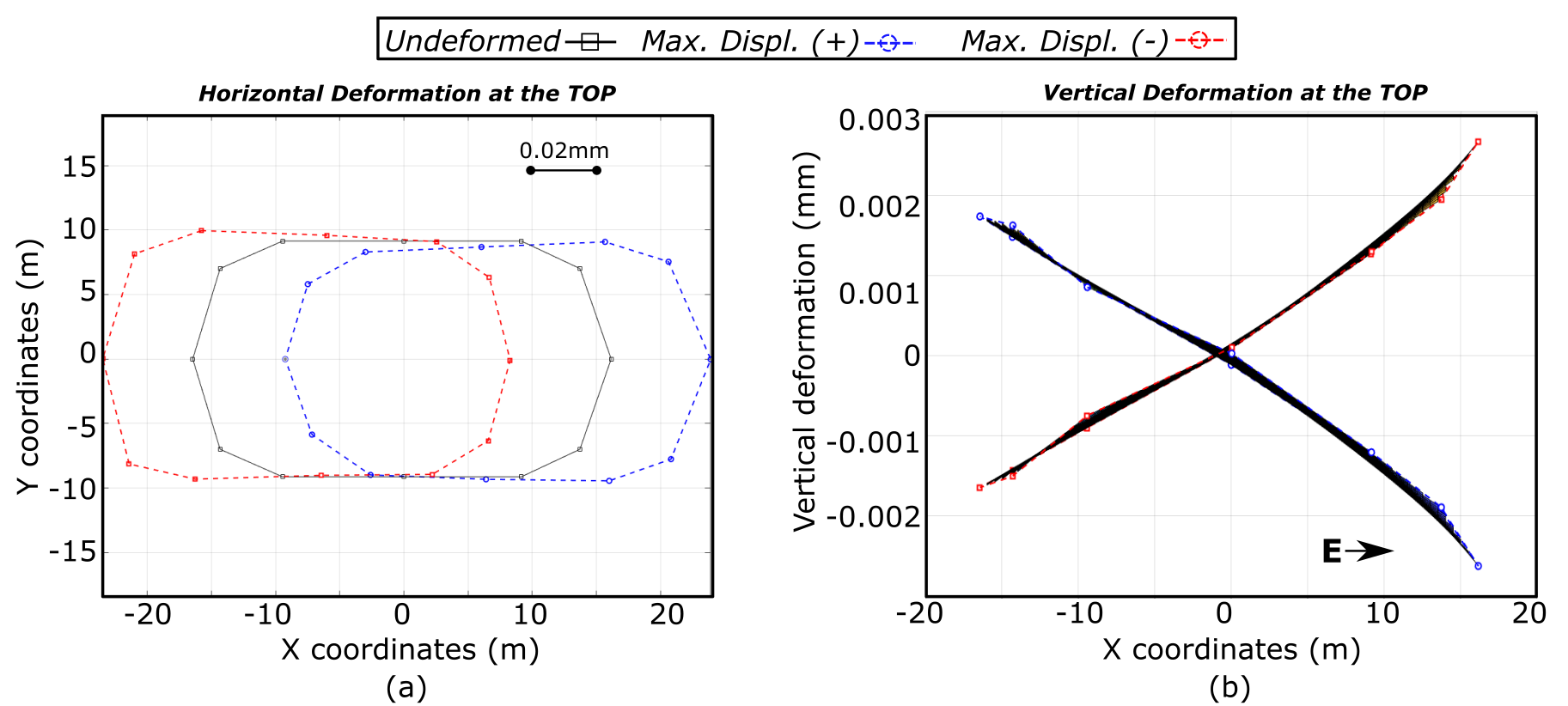} }
 \caption{Symmetry check of (a) total horizontal displacement and (b) vertical displacement at the top of the reaction mass }
  \label{fig3.9}
\end{figure}

\subsubsection{North-South Excitation (Y-DOF)}

The frequency response curves from excitation in the NS direction showed an increase in maximum displacement as the exciting frequency approached 10Hz, see Figures 3.10-3.12. Nonetheless, unlike the East-West results, the peak displacement was not as clear. Stations in groups T3, T2 which were located at the top corners of the reaction mass showed a maximum displacement at 9.5Hz, whereas groups T1, B2 which correspond to accelerometers in the top and bottom of the centerline of the structure respectively, show a not so clear peak between 12-14Hz.The difference in the response of different groups was also observed in the results from 2003, which indicated localized deformation towards the East-West walls \cite{Luco:2011}. Moreover, the vertical displacement amplitude increased in accordance with the horizontal response, suggesting the presence of rocking around the X-axis in a Y-translation-plus-roll mode of vibration. Given the localized response of some station, the rigid body frequency response curve, Figure 3.10(c) provided a better insight of the overall response of the reaction block. In this figure it was once again observed the presence of rocking in addition to the translation. Additionally, a broader peak displacement was identified between 9.5Hz and 12Hz agrees with the peaks observed in the different groups of accelerometers from Figure 3.10(a). Finally, the contribution of rigid body motion to the total displacement under the 10Hz excitation was 80\% and 50\% for the Y and Z translation respectively.

Regarding the deformation pattern shown in Figures 3.11,3.12, the localized deformation towards the east, west ends and the rocking behavior was observed as well. The deformation pattern was symmetric when looking at the maximum north and maximum south displacements, unlike the East-West response. Nonetheless, it could be seen that the deformation on the east side was slightly larger than that on the west. This can be attributed to the different soil types on each side as explained previously. Finally, it was observed that the horizontal response at the middle and bottom of the reaction mass was mostly rigid body motion, which was also observed in the East-west results. However, horizontal deformation at the top and vertical deformation was significant. In this case rigid body motion represented only 80\% and 50\% of the horizontal and vertical response respectively, showing less contribution of rigid body motion to the response.  Once again, the RBM contribution percentage correspond only to the results from the 10Hz excitation.

\begin{figure}[H]
 \centering
 {\includegraphics[width=.8\linewidth]{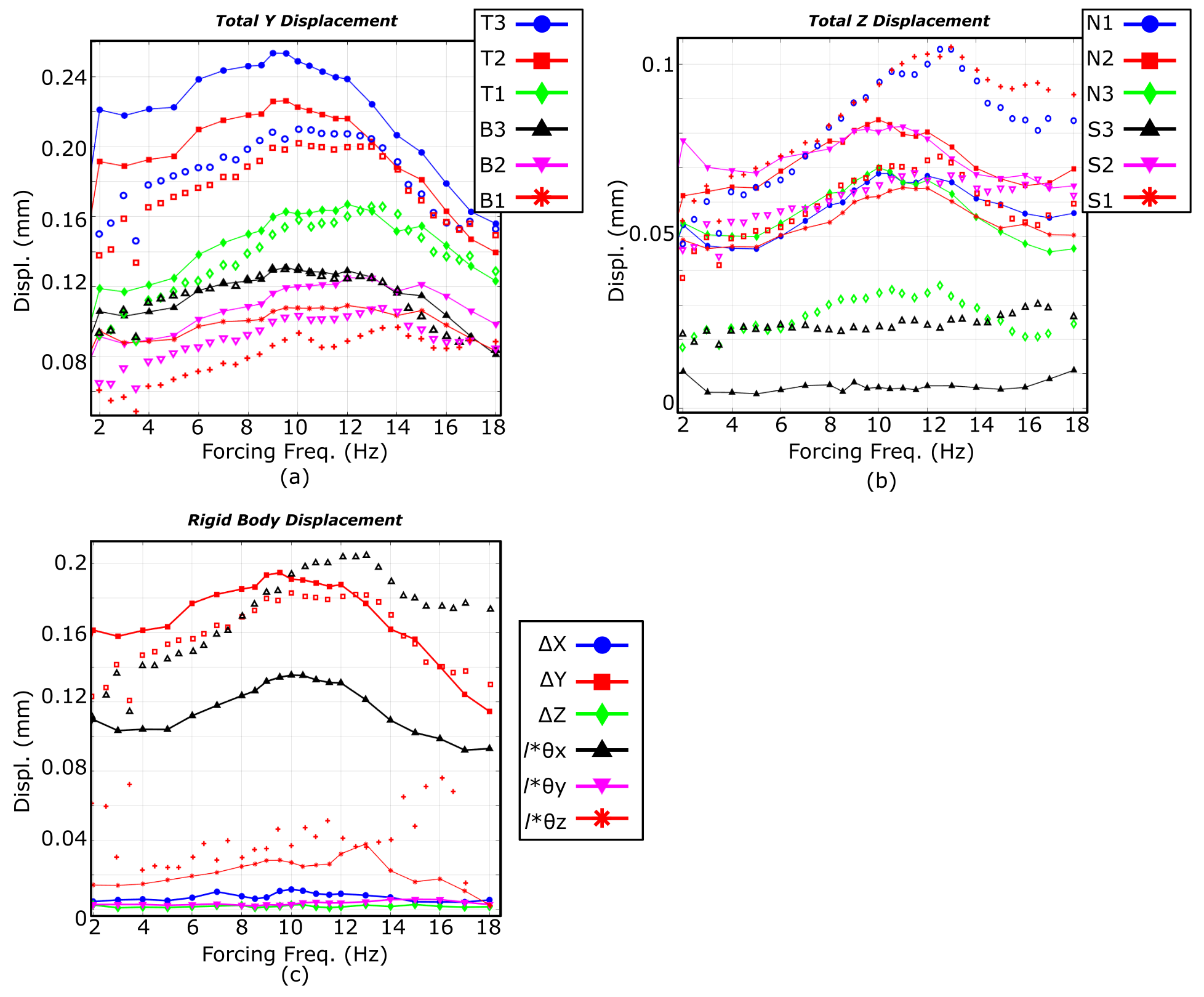} }
 \caption{Frequency Response Curve of North-South Excitation (Y-DOF). Comparison of Stepped Frequency tests (solid marker), 2003 tests (open marker): (a) Horizontal total displacement, (b) Vertical total displacement, (c) Rigid Body Horizontal Displacement.}
  \label{fig3.10}
\end{figure}

\begin{figure}[H]
 \centering
 {\includegraphics[width=.8\linewidth]{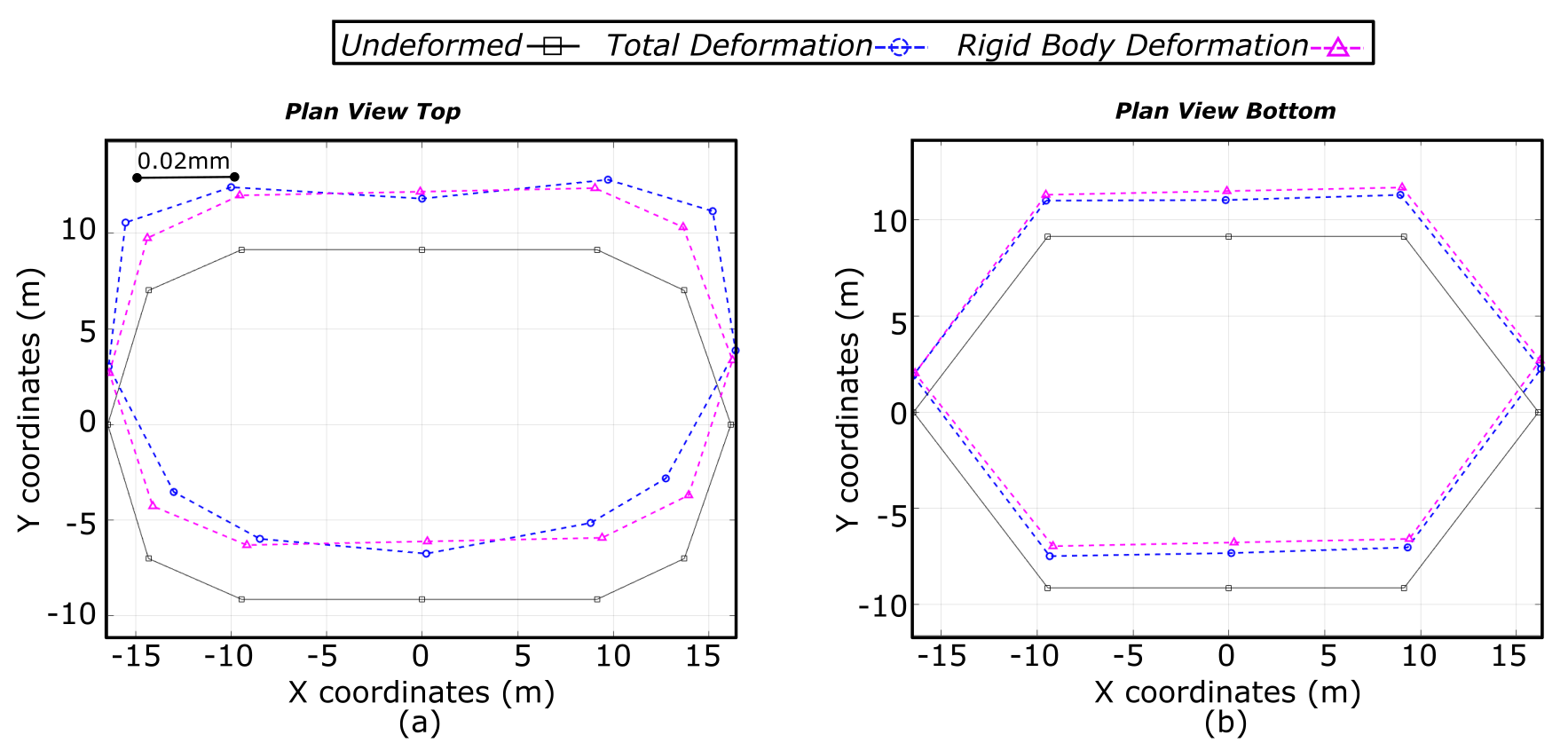} }
 \caption{Deformed configuration showing Total horizontal displacement (square) and rigid body displacement (triangle) from North-South Excitation (Y-DOF) at 10Hz: Top of reaction mass (a), bottom of reaction mass (b)}
  \label{fig3.11}
\end{figure}

\begin{figure}[H]
 \centering
 {\includegraphics[width=.8\linewidth]{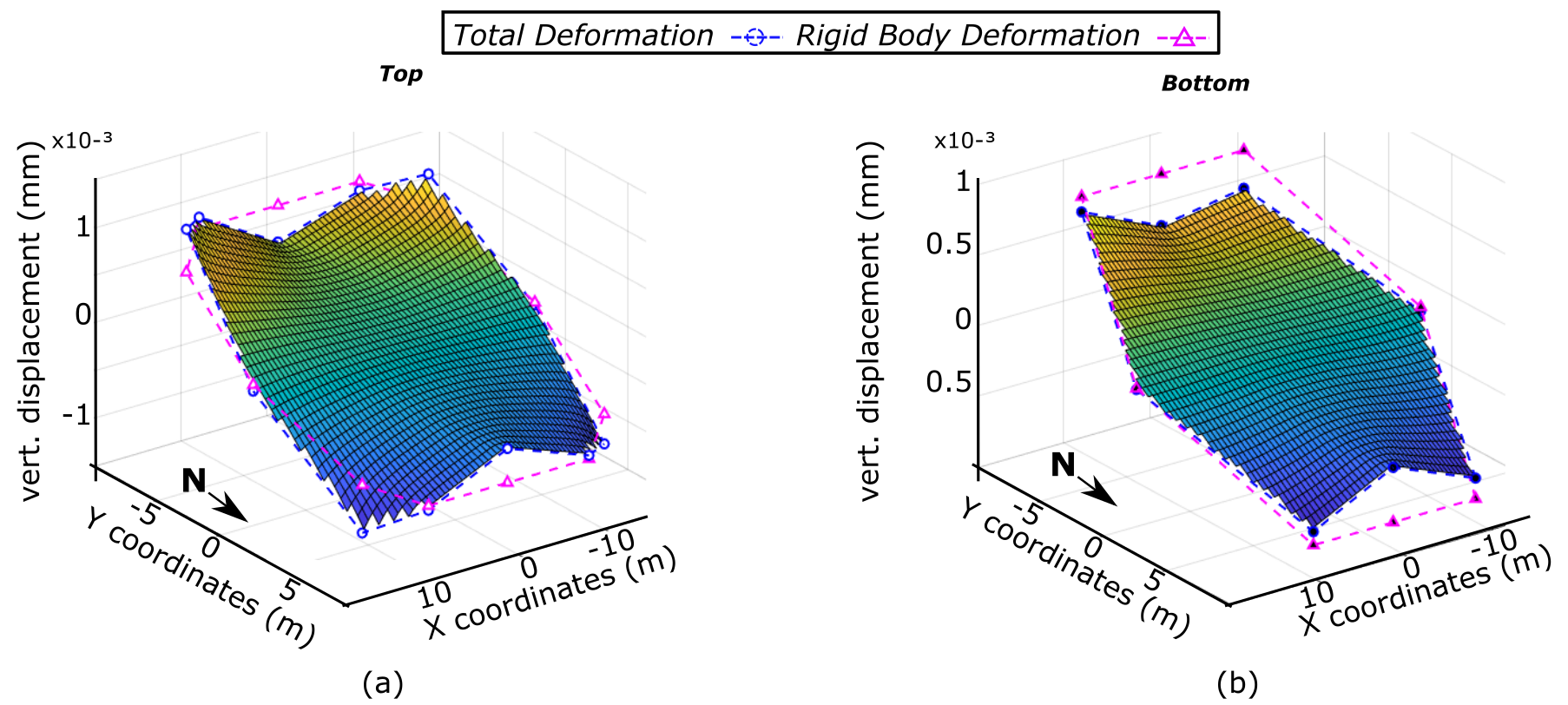} }
 \caption{Total vertical displacement (square) and rigid body vertical displacement (triangle) from North-South Excitation (Y-DOF) at 10Hz: Top of reaction mas (a), bottom of reaction mass (b)}
  \label{fig3.12}
\end{figure}

\subsubsection{Torsional Excitation (Yaw-DOF)}

During this set of excitations, the actuators were commanded to rotate the platen around the vertical axis. However, the forces onto the reaction mass were applied in the directions shown in Figure 2.1. Based on the resultant force, the corresponding torque was computed with the X,Y components of each actuator and the distance to the center point of the reaction mass. Moreover, to obtain the frequency response plots shown in Figures 3.13,3.14, the torque amplitude was linearly scaled to 117000 kN-m which corresponds to the maximum nominal actuator force of 6800kN multiplied by the distance between the shakers used during the 2003 tests. \citep{Luco:2011}. 

\begin{figure}[H]
 \centering
 {\includegraphics[width=.71\linewidth]{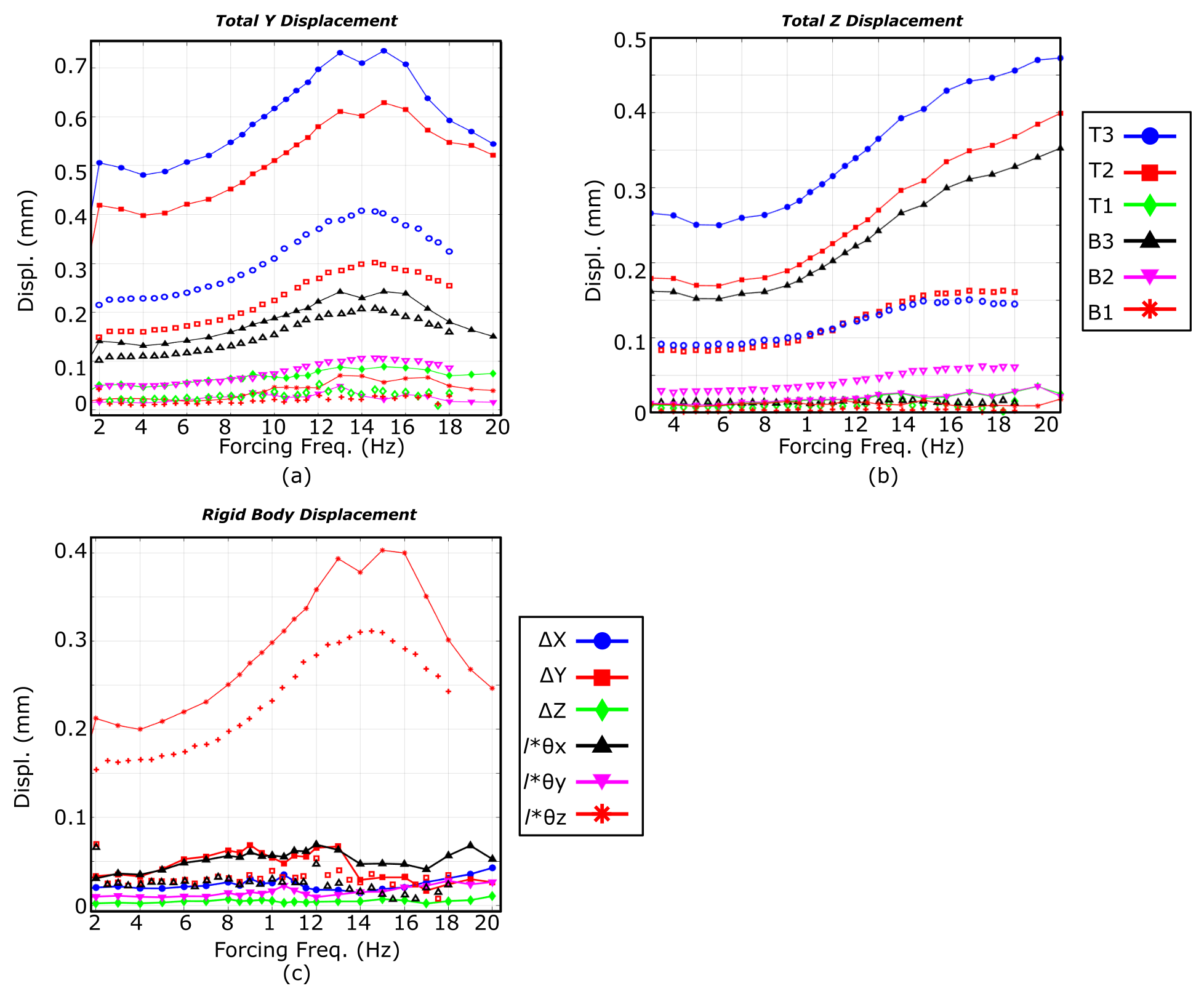} }
 \caption{Frequency Response Curve of Torsional Excitation (Yaw-DOF). Comparison of Stepped Frequency tests (solid marker), 2003 tests (open marker) : (a) Horizontal total displacement, (b) Vertical total displacement, (c) Rigid Body Horizontal Displacement. }
  \label{fig3.13}
\end{figure}

The horizontal displacement frequency response curve, Figure 3.13(a), peaked between 14Hz-15Hz. Unlike the North-south response, all groups of stations peaked at the same frequency. Nonetheless, the vertical response did not peak and kept increasing up to the highest excited frequency of 18Hz which was not observed in the two previous degrees of freedom. This continuous increase with no peak was not observed in the results from Luco (\citeyear{Luco:2011}) either.

Regarding the estimated rigid body response, the peak displacement could also be found at 14Hz-15Hz and most of the translation occurred in the North-south direction (Y-DOF). Unlike the previous two cases, the rigid body motion did not capture any rocking or motion in the Z vertical direction.

Figures 3.14,3.15 show the deformation pattern of the structure when excited at 14Hz. Like the response of previous degrees of freedom, horizontal displacement at the middle and bottom of the reaction mass is comprised mostly of rigid body motion. Displacement at the top on, the other hand presented a larger deformational component as seen in Figure 3.14(a). Finally, as previously mentioned, rigid body motion did not capture the vertical displacements, which is mostly deformational and as seen in the saddle-like pattern in Figure 3.15. This result was congruent with the computed contribution of rigid body motion determined as 60\% of the total horizontal displacement and only 15% of the total vertical displacement.

\begin{figure}[H]
 \centering
 {\includegraphics[width=.8\linewidth]{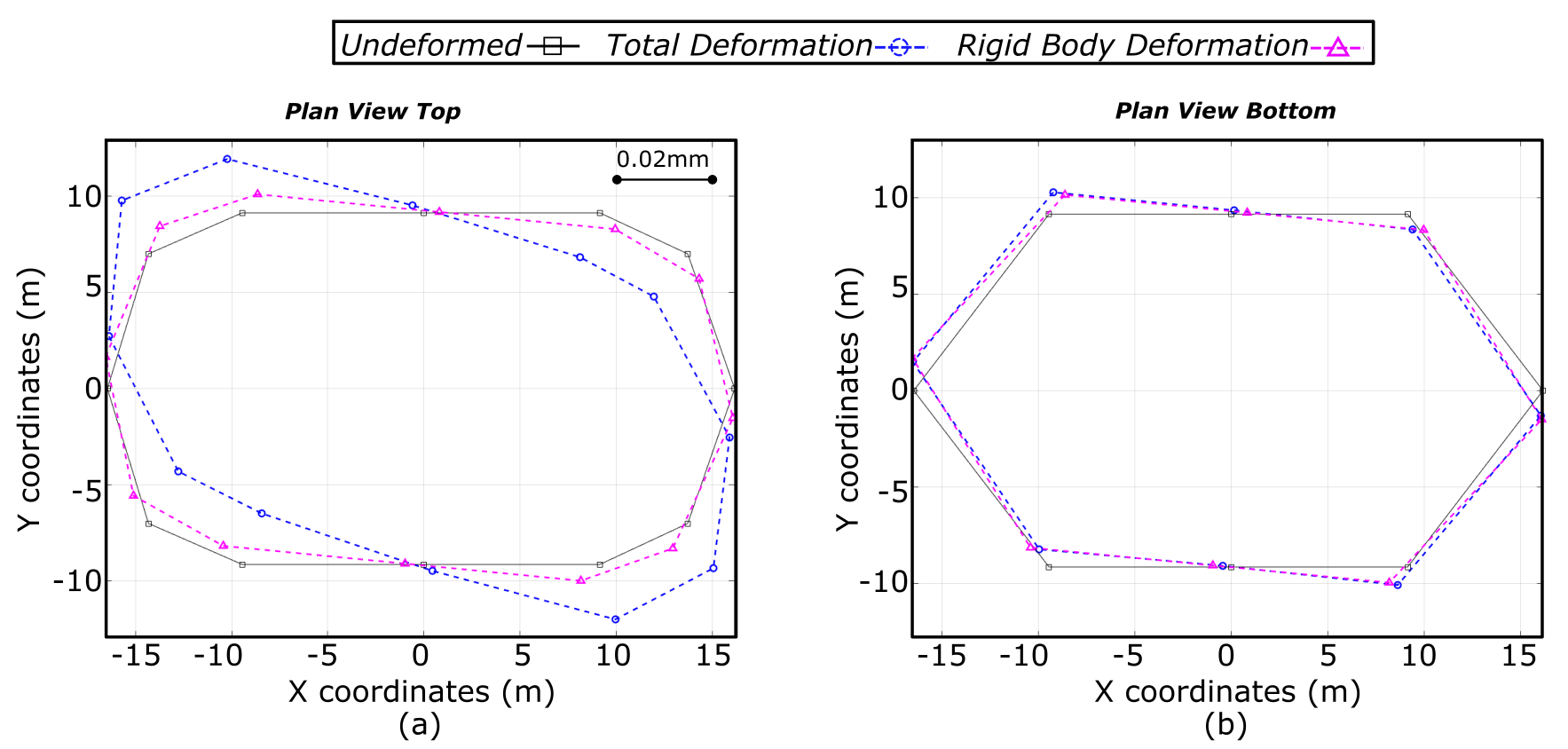} }
 \caption{Total horizontal total displacement (square) and rigid body displacement (triangle) from Torsional Excitation (Yaw-DOF) at 14Hz: Top of reaction mas (a), bottom of reaction mass (b)}
  \label{fig3.14}
\end{figure}

\begin{figure}[H]
 \centering
 {\includegraphics[width=.8\linewidth]{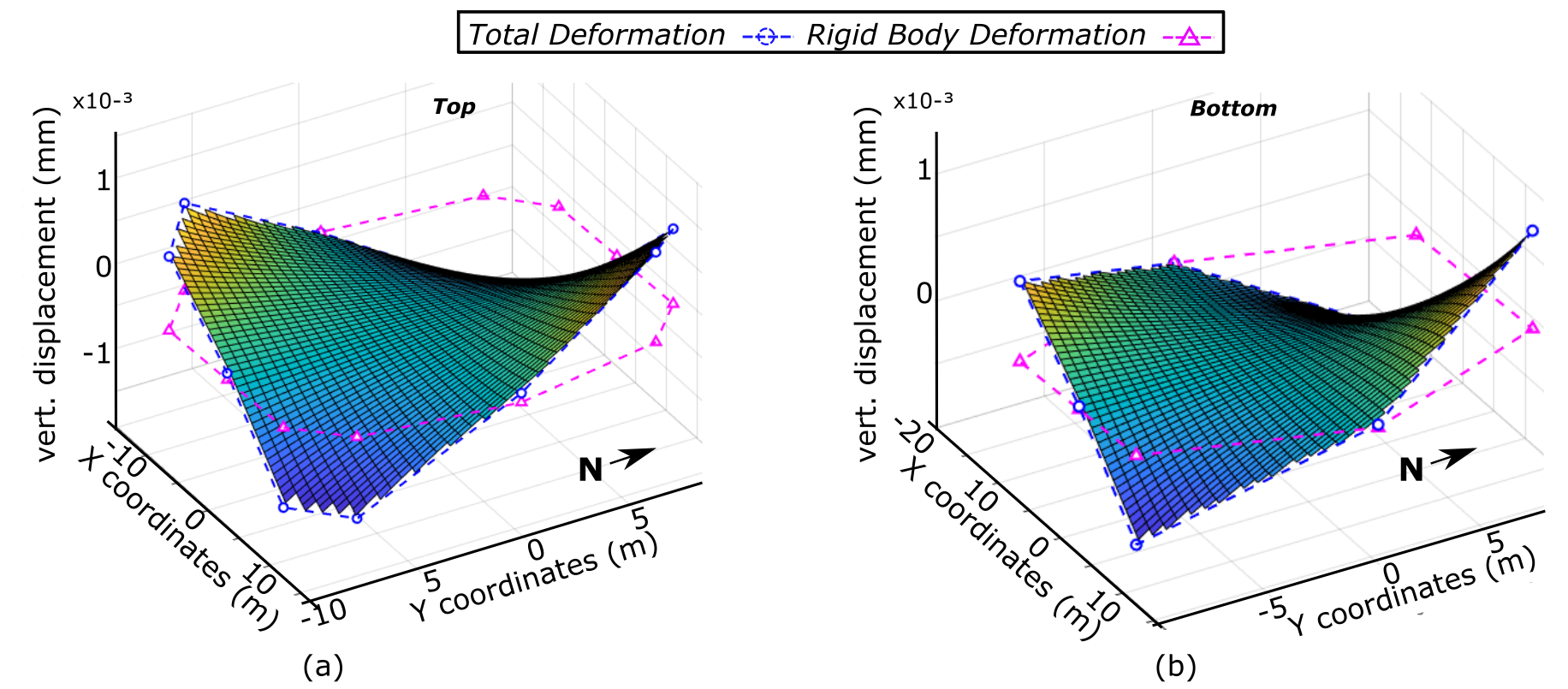} }
 \caption{Total vertical displacement (square) and rigid body vertical displacement (triangle) from Torsional Excitation (Yaw-DOF) at 14Hz: Top of reaction mas (a), bottom of reaction mass (b)}
  \label{fig3.15}
\end{figure}

\subsubsection{Vertical Excitation (Z-DOF)}

This degree of freedom, not tested previously, was excited with the six new vertical actuators located underneath the platen in a range of frequencies between 5Hz and 25Hz. Like previous tests, displacements were scaled to the same amplitude of 6800kN although this was not the nominal maximum force of the vertical actuators. 

Frequency response curves of this DOF showed a maximum displacement occurring between 15Hz-16Hz, except the station labeled as B3 as seen in Figure 3.5. This station corresponds to the accelerometer placed close to the center bottom of the reaction mass near the Hold Down struts shown in Figure 2.3. This location was further away from the surrounding walls and right next to a point of applied force; thus, a larger displacements and localized behavior was expected. Moreover, the rigid body frequency response curve, see Figure 3.16(b), shows a peak between 12Hz-16Hz and it was observed that this mode of vibration was mostly translational with little to no contribution of rocking in any direction.

When inspecting the deformation pattern, it was observed that the top of the reaction mass presented some horizontal deformation, while the motion at the middle and bottom was entirely vertical as shown in Figures 3.17(a) and (b) respectively.  Finally, from the vertical displacement shown in Figure 3.18, the stations along the centerline of the reaction mass presented a larger deformation than at the corners, which were expected to be more rigid. This additional deformation was not accurately captured by the estimated rigid body displacement. Consequently, the contribution of rigid body motion represented only 78\% of the total displacement in the vertical direction. 

\begin{figure}[H]
 \centering
 {\includegraphics[width=.9\linewidth]{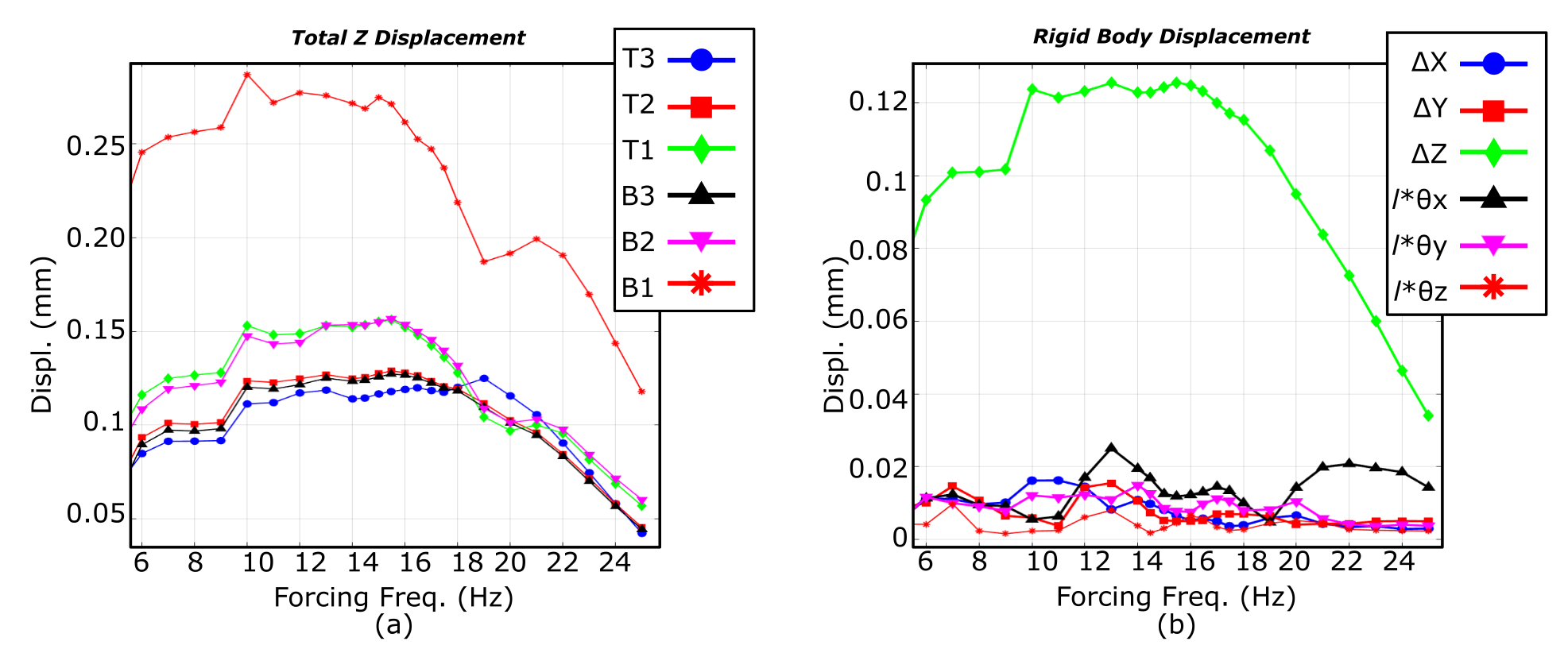} }
 \caption{Frequency Response Curve of vertical excitation (Z-DOF), (a) Vertical total displacement, (b) Rigid Body Vertical Displacement. }
  \label{fig3.16}
\end{figure}

The total deformation of the south face, displayed in Figure 3.17 (c) resembled a beam subjected to bending. Using this elastic beam assumption, where plane sections remain plane, an approximation of the curvature was calculated to better understand the magnitude of the deformations. First, the reaction mass deformation was computed by subtracting the Rigid body displacement from the total. Subsequently, a parabola was fitted using the 3 stations at the top of the face shown in Figure 3.17(c). Finally, the curvature was approximated as the second derivative of such parabola. With the curvature and the elastic beam assumptions, a tensile strain of $1.78x10^{-6}$ was determined, which corresponded to less than 5\% of the tensile strength of concrete. The tensile strength for a 41MPa concrete was calculated using Vecchio and Collins (\citeyear{vecchio:1986}) strain-stress relationship, and ACI318-19 \citep{ACI:2019} for the modulus of elasticity. Given the assumptions just mentioned, the computed strain from experimental results and the tensile strength are just approximations and can only be considered upper bound limit of the deformations that the reaction mass can sustain.

\begin{figure}[H]
 \centering
 {\includegraphics[width=.8\linewidth]{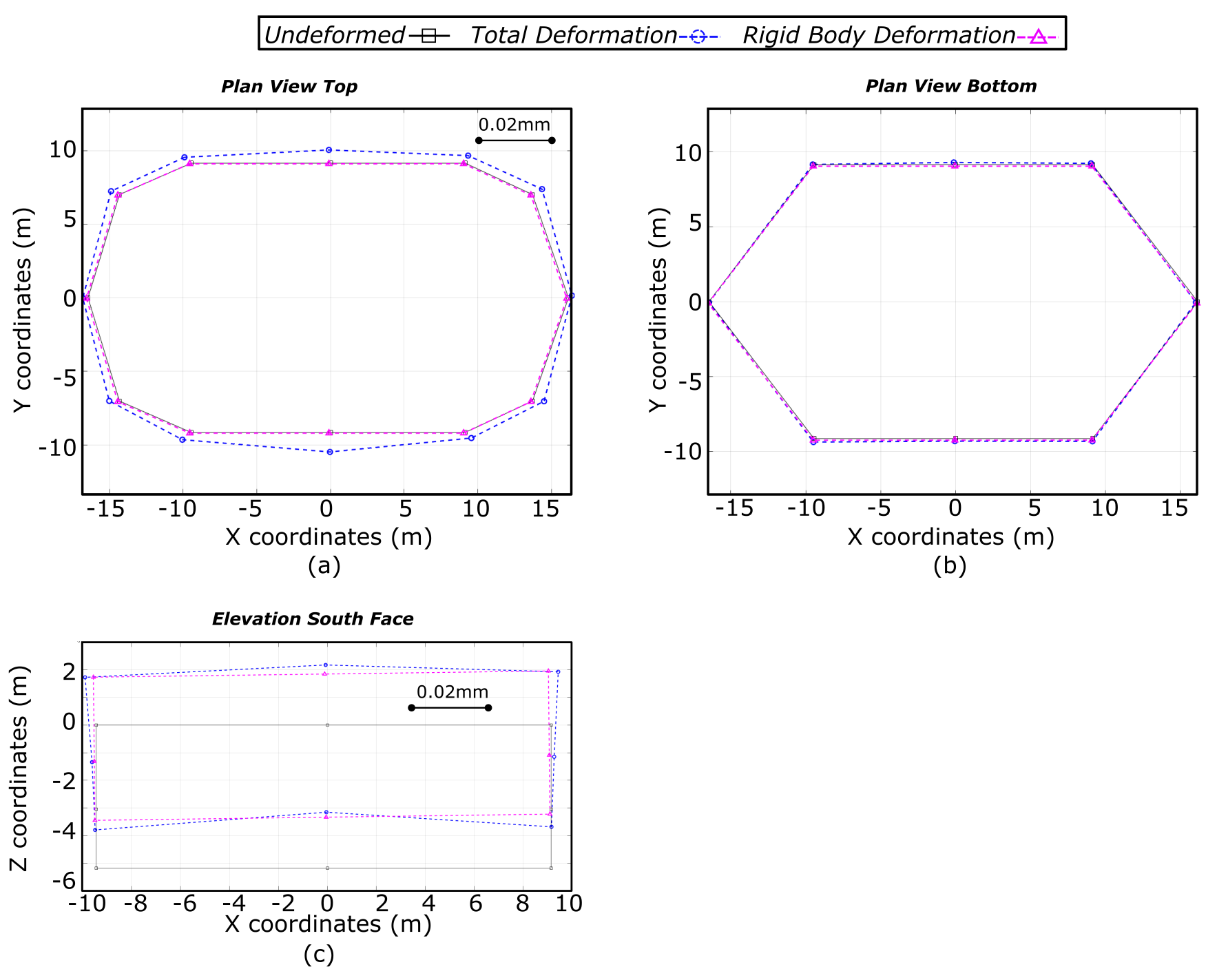} }
 \caption{Deformed configuration showing Total horizontal displacement (square) and rigid body displacement (triangle) from vertical Excitation (Z-DOF) at 14Hz: Top of reaction mass (a), bottom of reaction mass (b), South face elevation (c) }
  \label{fig3.17}
\end{figure}

\begin{figure}[H]
 \centering
 {\includegraphics[width=.8\linewidth]{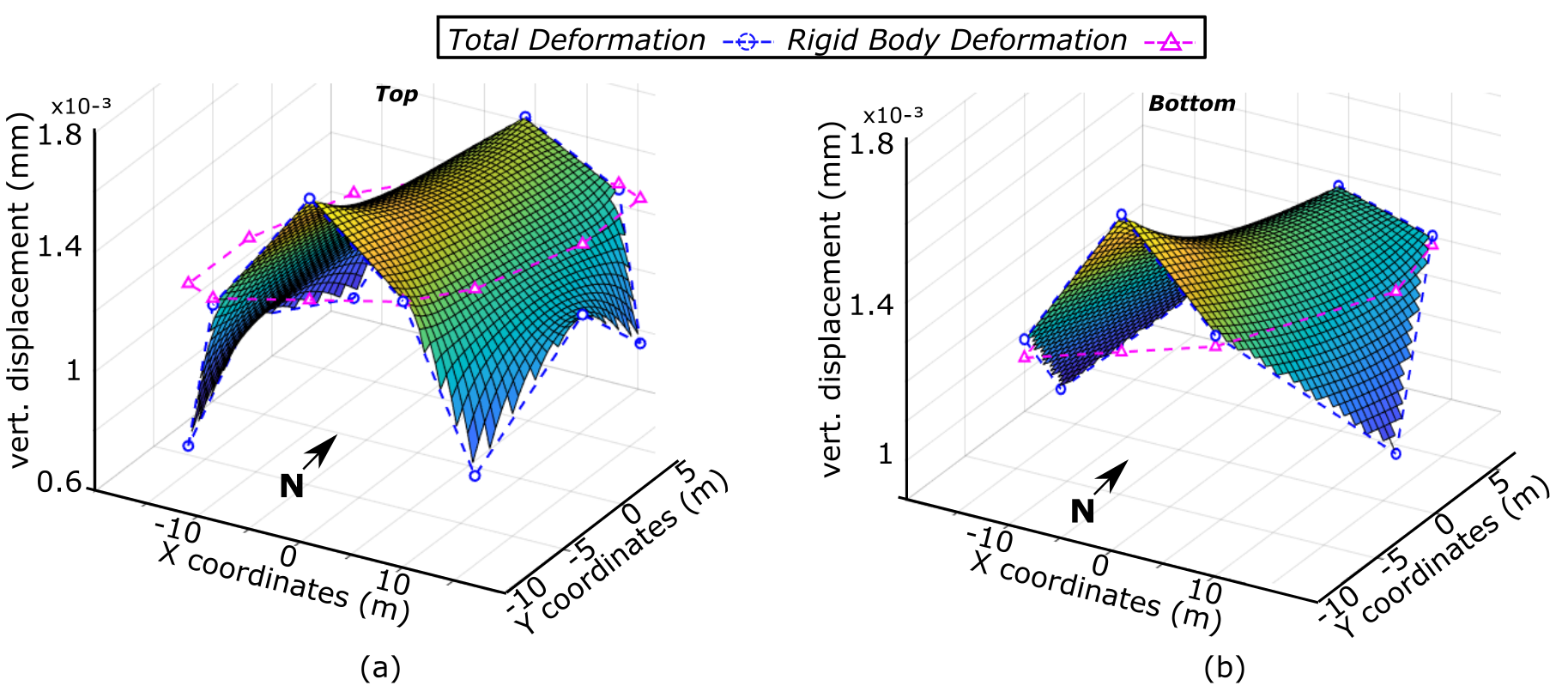} }
 \caption{Total vertical displacement (square) and rigid body vertical displacement (triangle) from vertical excitation (Yaw-DOF) at 15Hz: Top of reaction mas (a), bottom of reaction mass (b)}
  \label{fig3.18}
\end{figure}

\subsection{Sweep Frequency Results}

Sweep frequency response data was used to generate frequency response plots like those obtained in the previous section. Given that Sweep Frequency Tests covered all frequencies of interest in a single excitation, it was possible to repeat the same test at different force amplitudes as described on Table 2.1. The repetition of the same test at a larger force allowed to evaluate the accuracy of the scaling of displacements from one test to the next one in a procedure referred as Test of Linearity discussed in the upcoming section. Finally, the rigid body frequency response was compared to that of the stepped frequency tests to gain confidence in the calculated maximum displacements obtained with both stepped and sweep frequency tests.

Figure 3.19 shows the rigid body frequency response curve of each degree of freedom excited: x, y, yaw, z respectively, all scaled to a force amplitude of 6800kN for the translational degrees of freedom and a torque of 117000 kN-m for the Yaw excitation. At first glance, all tests in all degrees of freedom tested showed good agreement on the maximum scaled displacement for all frequencies. The response to East-West excitation displayed the best agreement on all frequencies covered. Results from the North-South and torsional excitation showed good agreement as well, but only at higher frequencies. Results from vertical excitation on the other hand, presented an increase in displacement past 20Hz which was not observed on the steeped frequency results. This is further discussed in section 4. Finally, to increase the confidence the linearly scaled results, the Root Mean Squared difference (RMS) was determined among sweep test results and between sweep and stepped test results, see Table 3.1. 

\begin{figure}[H]
 \centering
 {\includegraphics[width=.8\linewidth]{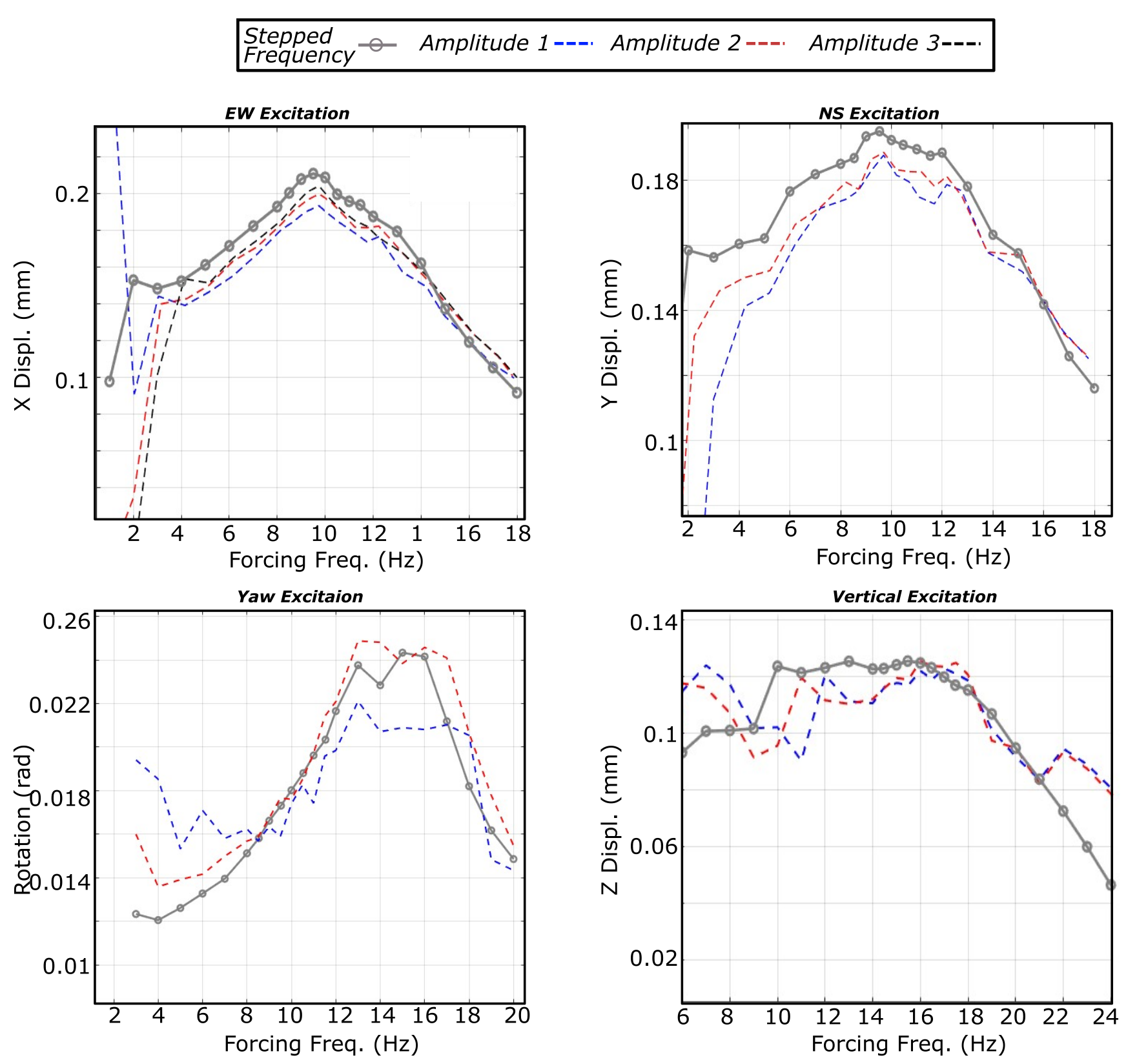} }
 \caption{Comparison of Rigid Body Frequency response curve in the X, Y, Yaw and Z degree of freedom respectively between steeped tests (gray) and sweep tests (red and blue), all scaled to the same force/torque amplitude.}
  \label{fig3.19}
\end{figure}

\subsection{Effective Damping Ratio of the System}

Using the frequency response curves it was possible to estimate the dynamic amplification factors and damping ratio for each degree of freedom tested. The dynamic amplification factor corresponds to the ratio between the peak dynamic and static response, in this case displacement. Determining the static response experimentally using a vibration generator can be difficult (Chopra 2002), therefore, the response at very low frequencies was the best approximation available to the static displacement and used for the amplification factors summarized on Table 3.1. Furthermore, the experimental frequency response curves were normalized and compared to the response of a single degree of freedom oscillator as a first-order estimation of the damping ratio. As described by Chopra (2002), the normalized frequency response curve of the oscillator subjected to a harmonic loading is defined by:

\begin{equation}
    \ddot{u}+2\xi \omega_n\dot{u}+\omega_n^2 u= \frac{P_o}{m} sin(\omega\ t) 
\end{equation}
\begin{equation}
    R_d= \frac{u_{static}}{u_{dynamic}} = 1 / 
         \sqrt{\left(1-\left(\frac{\omega}{\omega_n}\right)^2\right)^2 + \left( 2\xi\left(\frac{\omega}{\omega_n}\right)^2\right)} 
\end{equation}

Where (3.5) is the equation of motion of the SDOF oscillator and (3.6) is the transfer function defining the dynamic amplification factor.

Figure 3.20 shows $R_d=f(\omega,\xi)$ plotted on top of the normalized frequency response curves from the East-West excitation. Considering all 29 accelerometers shown in the figure, the damping ratio was estimated between 35\% and 45\% for this mode of vibration as shown in Figure 3.20(a). These results were consistent with the results presented by Luco (\citeyear{Luco:2011}) which estimated an overall damping ratio between 32\% and 42\%. Results for the other degrees of freedom tested are summarized on Table 3.1.

\begin{figure}[H]
 \centering
 {\includegraphics[width=.8\linewidth]{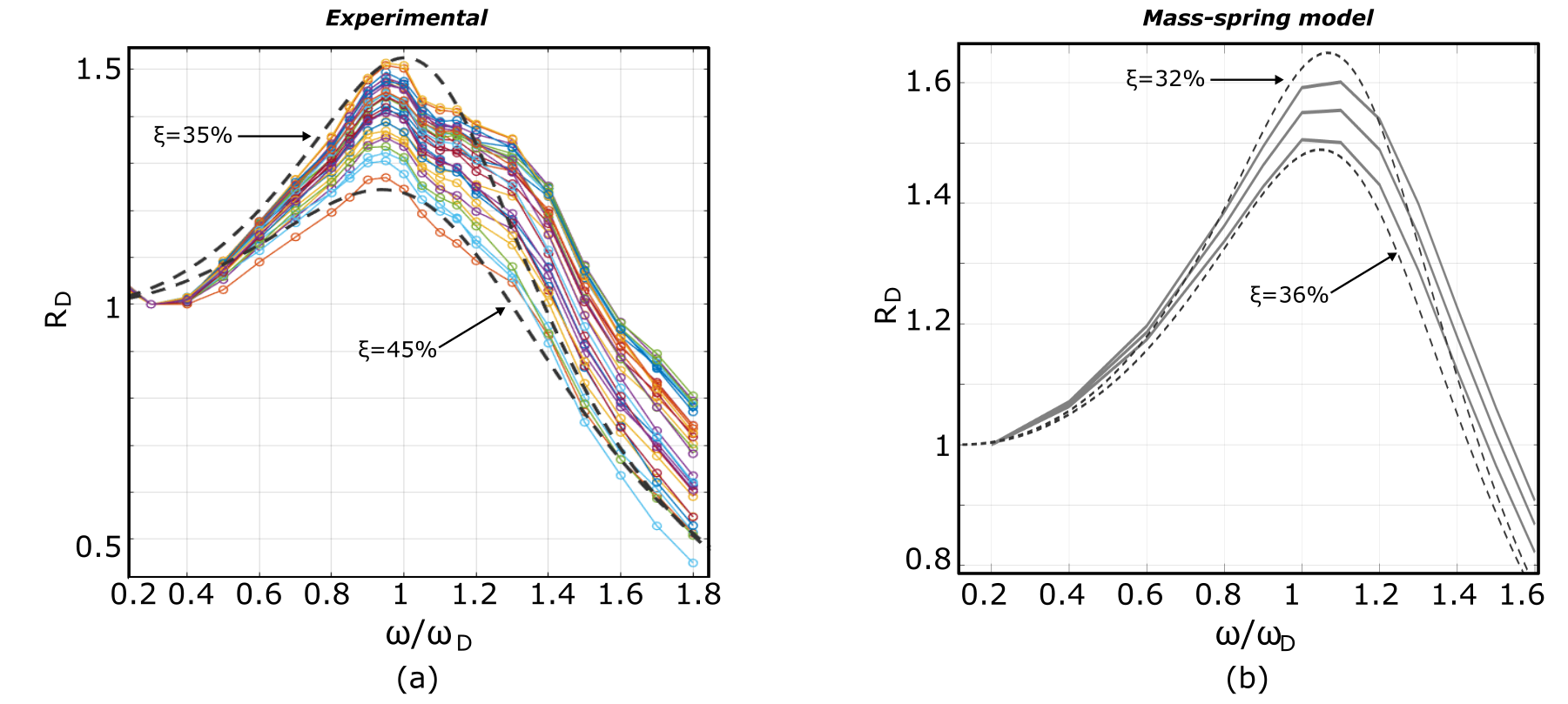} }
 \caption{Dynamic amplification factors from the East-west excitation compared to dynamic amplification of a SDOF oscillator with different damping ratios. (a) Experimental results, (b) synthetic data}
  \label{fig3.20}
\end{figure}

As an exercise, the damping ratio estimation procedure was also applied on the synthetic data, see Figure 3.20(b). The estimated damping ratio between 32\% and 33\% was close to the 37\% ratio assigned to the dashpots in the model.

\begin{table}[H]
\caption{Dynamic Properties obtained from Force vibration tests}
 \label{table:assembly3}
\centering
%\small
\begin{tabular}{
>{\centering\arraybackslash}m{5em}|
>{\centering\arraybackslash}m{3em}|
>{\centering\arraybackslash}m{3em}|
>{\centering\arraybackslash}m{3em}|
>{\centering\arraybackslash}m{3em}|
>{\centering\arraybackslash}m{3em}|
>{\centering\arraybackslash}m{3em}|
>{\centering\arraybackslash}m{2em}|
>{\centering\arraybackslash}m{2em}|
>{\centering\arraybackslash}m{6em} } 
\hline\hline
{\textbf{Vibration Mode}} & 
\multicolumn{2}{|m{6em}|}{\textbf{Natural Frequency [Hz]}} & 
\multicolumn{2}{|m{6em}|}{\textbf{Damping Ratio \%}} & 
\multicolumn{2}{|m{6em}|}{\textbf{Peak Scaled Displ. (mm)}} &
\multicolumn{2}{|m{4em}|}{\textbf{Peak amplification}} &
{\textbf{RBM Contribution \%}} \\
\hline
&\textbf{2022}&\textbf{2003}&\textbf{2022}&\textbf{2003}&\textbf{2022}&\textbf{2003}& \textbf{2022}&\textbf{2003}& \\
\hline
 X+pitch & 9.5 & 10 & [35-45] & [32-42] & 0.25 & 0.26 & 1.33 & 1.30 & X=94 Z=89\\
 Y+roll  & 9.5 & 10 & [35-50] & [32-42] & 0.25 & 0.21 & 1.26 & 1.40 & Y=80 Z=50\\
 Yaw     & 14  & 14 & [25-40] & [32-42] & 0.75 & 0.4  & 1.64 & 1.80 & Y=60 Z=15\\
  Z      & 15  & -  & [35-50] &    -    & 0.27 &  -   & 1.62 &   -  & Z=78\\

\hline
\end{tabular}
\normalsize
\end{table}

\section{ANALYSIS AND DISCUSSION OF RESULTS}

Considering that all displacements were linearly scaled to a specified force/torque to compare across experiments, the first subsection focuses on the linearity of the system. Furthermore, a qualitative comparison between the most recent set of tests and the ones performed in 2003 is presented. The comparison discusses the possible sources of discrepancy and the possible changes in the reaction mass-soil system after the 6-DOF upgrade. Finally, this section examines the contribution of rigid body motion to total displacement, whose results were previously presented in Table 3.1. 

\subsection{Test of Linearity}

The agreement between frequency response curves from tests at different amplitudes provided an insight on the level of linearity of the reaction mass-soil system. In perfect linear behavior, displacements increase proportionally with the force applied. A certain degree of linearity was expected in the non-linear system given the very small displacements, which were several orders of magnitude smaller than the size of the structure, and the applied forces which generated stresses of less than 1\% the bearing capacity of the soil.  Consequently, the scaled results represented only an approximation of the displacements expected if the maximum force of 6800kN were to be applied. The quality of this approximation depended on the magnitude of the scaling factor and the level of non-linearity of the system. A crucial aspect to consider was that none of the most recent tests applied a force nearly close to the benchmark of 6800kN. Therefore, the results discussed only represented a relative linearity between sweep and stepped frequency test.

To quantitatively evaluate the linearity, sweep frequency results were analyzed since the same test was repeated with the only variable being the applied force. At first sight the agreement between different sweep tests, shown as dashed curves in Figure 3.19, suggested a considerable degree of linearity. A closer look at the Root Mean Square (RMS) of the difference between the results corroborated the visual inspection. It should be noted that results from excitation at 1Hz and 2Hz were excluded when computing the root mean squared difference since they were not representative of the response. Values from Table 4.1 show than when comparing only the sweep test, there was an overall maximum mean square difference of 0.01mm or 5\% of the maximum displacement. When looking at the X-DOF response, Figure 3.19(a) the discrepancies were even smaller suggesting a greater degree of linearity for this mode of vibration.  
Moreover, rigid body motion was compared between sweep frequency and stepped frequency tests. In this case, differences were larger than before, see Table 4.1, with an overall maximum RMS difference of 0.017mm or 8\% of the maximum displacement. Nonetheless, more variables came into play when comparing between these different types of tests. The main difference being the excitation data and how it affected the calculation of the force amplitude used for the scaling factors. In the case of the stepped frequency tests, a long duration steady state response was recorded for each frequency of interest which made it straightforward to determine the force amplitude via least square fit.  On the other hand, during the sweep test, the frequency of interest was only maintained constant during a brief period before increasing, providing less data for the least square fit, especially at low frequencies. The comparison of the actuator force from the different types of tests and the amplitude found via LSF is presented in Figure 4.1. In addition, given the continuous change in frequency and the actuators being commanded to only maintain constant acceleration, the actuator force during these short windows was not perfectly constant, introducing more variability into the force amplitude used for the scaling factor as seen in Figure 4.1(b). 

The effects of the scaling factor can be observed in the discrepancies between tests in Figure 3.19, especially at low frequencies. Results from the Sweep test in the Yaw-DOF at the lowest force amplitude, Figure 3.19(c) displayed the largest discrepancies overall. A closer look at this test showed that neither the applied force nor the accelerometer response showed clear constant amplitude during the window of time frequency was maintained constant. This resulted in difficulty and uncertainty when determining the displacement and force amplitude used for scaling, influencing the final displacement results. Nevertheless, the overall frequency response curve and the peak displacements found from different tests still suggests a considerable degree of linearity, particularly in the X and Y-DOF which, despite the additional variables, the still maintained a remarkable agreement between tests. 

\begin{table}[H]
\caption{Test of linearity results.}
 \label{table:assembly4}
\centering
%\small
\begin{tabular}{
>{\centering\arraybackslash}m{6em}
>{\centering\arraybackslash}m{9em}
>{\centering\arraybackslash}m{10em}
>{\centering\arraybackslash}m{6em}
>{\centering\arraybackslash}m{7em} } 

\hline\hline
{\textbf{Test}} & 
{\textbf{Force Amplitude range (kN)}} & 
{\textbf{Avg. Amplification factor}} & 
{\textbf{Sweep test RMS (mm)}} &
{\textbf{stepped Test RMS (mm)}} \\
\hline

X Sweep 1   & [325-464]   & 22.2 & -     & 0.012 \\
X Sweep 2   & [915-1063]  & 6.8  & 0.008 & 0.013 \\
X Sweep 3   & [1696-1880] & 5.0  & 0.01  & 0.013 \\
\hline
Y Sweep 1   & [368-544]   & 22.2 & -     & 0.013 \\
Y Sweep 2   & [852-1160]  & 10.8 & 0.008 & 0.011 \\
\hline
Z Sweep 1   & [120-156]   & 19.6 & -     & 0.017 \\
Z Sweep 2   & [812-1160]  & 9.1  & 0.008 & 0.016 \\
\hline
Yaw Sweep 1 & [177-490]   & 20.5 & -     & 0.002 \\
Yaw Sweep 2 & [390-135]   & 38.4 & 0.002 & 0.001 \\

\hline
\end{tabular}
\normalsize
\end{table}

A particular response was observed on the actuator force applied during the Yaw-DOF sweep frequency test. During the entire excitation, a low frequency wave of 0.5Hz prevailed on top of the forcing frequency of interest, as seen in Figure 4.2. To accurately obtain the amplitude of the forcing frequency of interest, the low frequency wave was subtracted from the initial signal and the LSF procedure was once again applied. 

Finally, to better understand the extent of the sources of discrepancy just mentioned, the same stepped vs. sweep analysis was applied to the synthetic data from the simplified rigid model. In this case, applied forces had a constant amplitude of 6800kN for both types of tests, eliminating the need of any signal filtering, and most importantly scaling. In this case the differences between frequency response curves from different tests were almost indistinguishable suggesting that the variability of the force amplitude on the experimental tests had a considerable influence in the discrepancy of the scaled of displacements. 

\begin{figure}[H]
 \centering
 {\includegraphics[width=.9\linewidth]{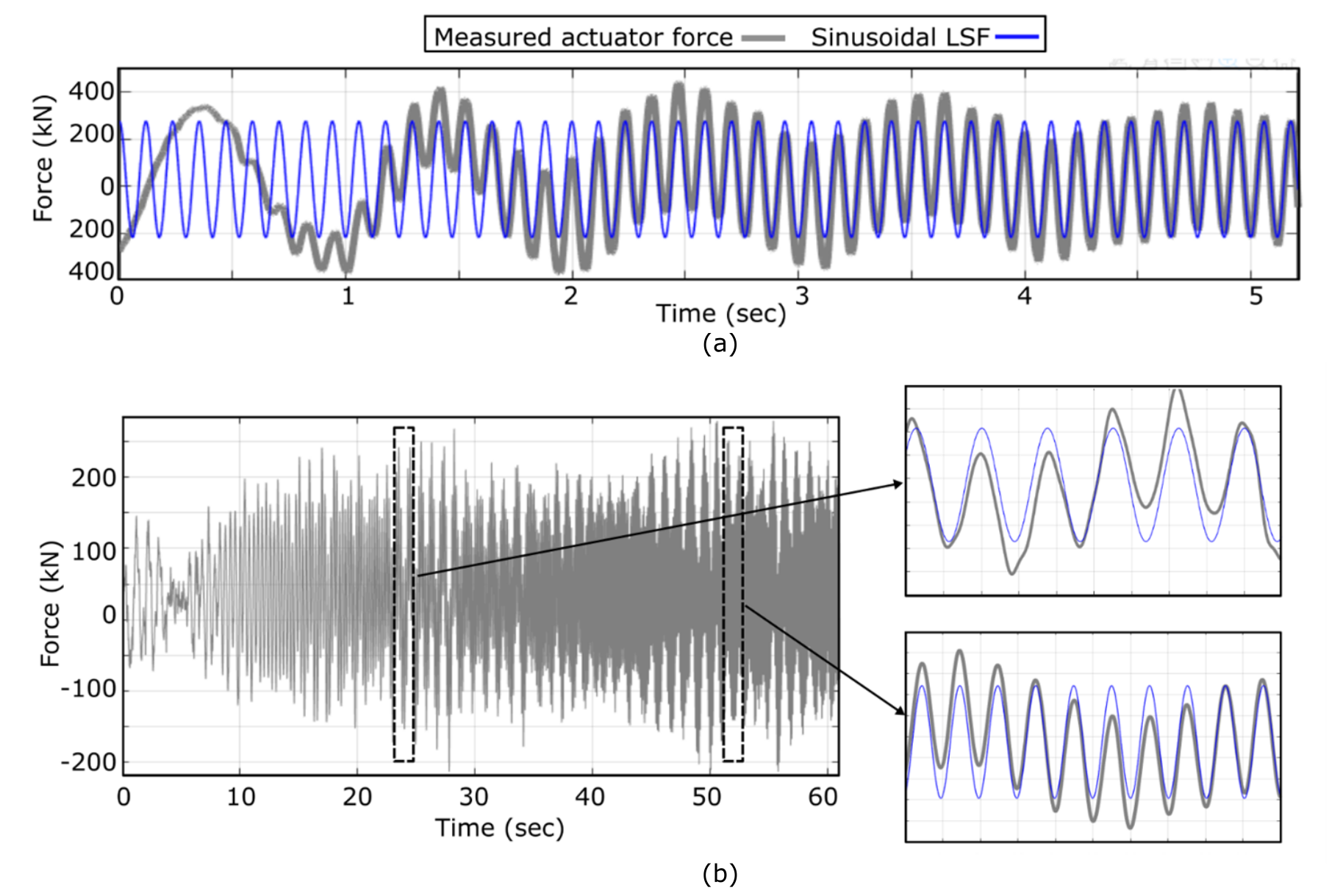} }
 \caption{Sample of least square fit used to determine force amplitude for the scaling factor. Comparison between (a) Stepped frequency NS excitation at 8.5Hz and (b) Sweep frequency NS excitation.}
  \label{fig4.1}
\end{figure}

\begin{figure}[H]
 \centering
 {\includegraphics[width=.7\linewidth]{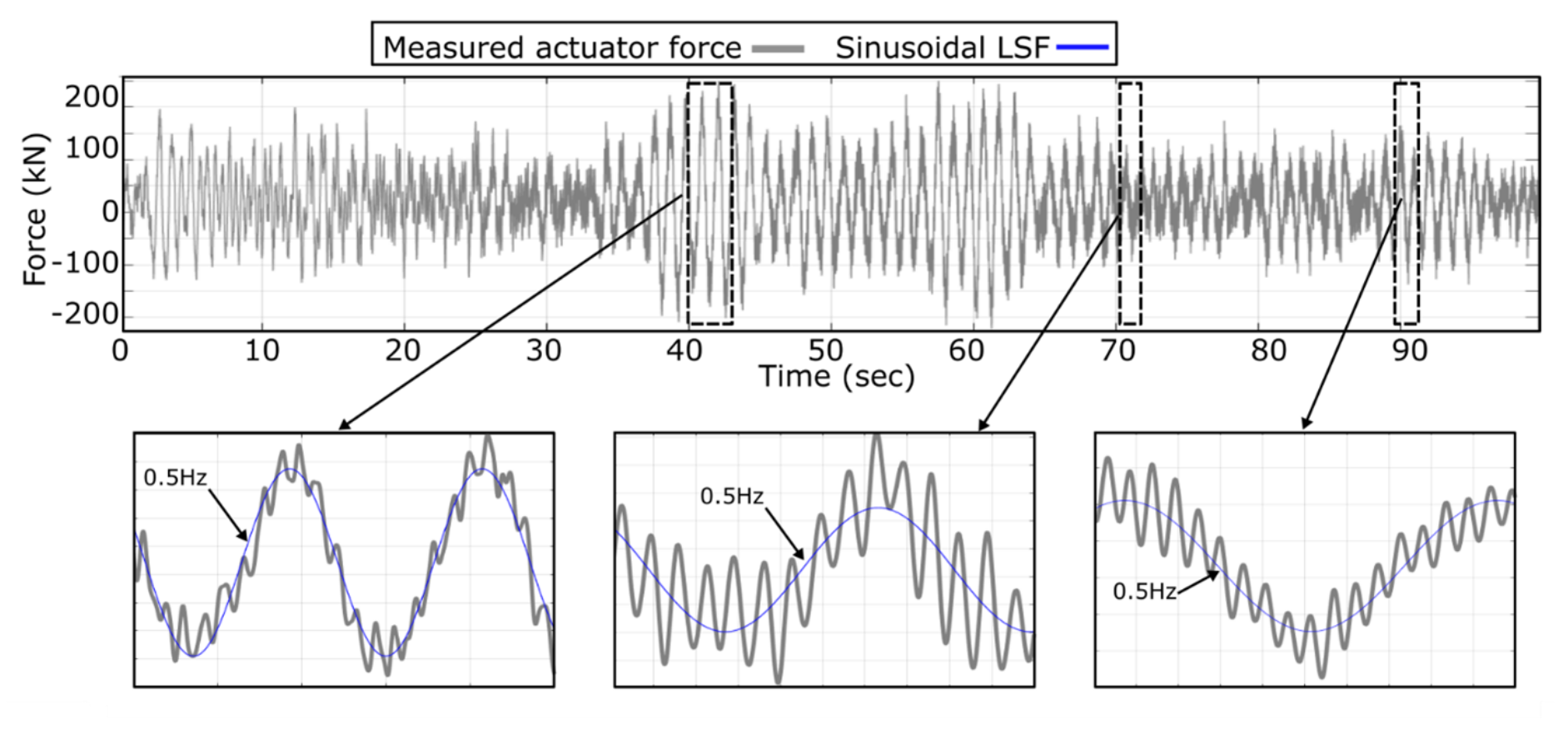} }
 \caption{Sample of least square fit with presence of low frequency waves on the recorded actuator force during the Yaw Excitation sweep frequency test}
  \label{fig4.2}
\end{figure}

\subsection{2022 Post-Upgrade Tests vs. 2003 Tests}

\subsubsection{General comparison of Results}

When looking at the frequency response curves, the 2022 results showed overall larger total displacements in all degrees of freedom compared to past tests, except for the group of top corner accelerometers, T3, in the X-DOF, Figure 3.6(a). In addition, larger differences were observed in the total displacements from the Yaw excitation which almost double the results from 2003, as seen in Figure 3.13. This specific case suggested that the degree of non-linearity of the Yaw mode of vibration was greater than in other DOF’s and the linear scaling approximation might not be as good. The increased level of non-linearity was also consistent with the very little contribution of Rigid Body motion that was observed in this response. Other than the nonlinearity of the system, additional sources of discrepancy contributed to the differences in the results and are discussed in the upcoming subsection. 

Aside from the differences in magnitude, the frequency response curves from both tests showed very similar behavior, suggesting that the dynamic properties of the reaction mass-soil system have not changed significantly over the years. Nonetheless it was observed that peak displacements occurred at frequencies within a range of 0.5Hz to 1Hz. This slight difference on the natural frequency can be difficult to characterize and cannot be interpreted as a change on the structure’s stiffness, as it could be just a byproduct of data processing or a combined effect of the sources of discrepancy later discussed. For example, the Millikan library in Pasadena, California has been extensively studied under forced vibration tests over the decades \citep{Kuroiwa:1967,Foutch:1976,Chopra:2002,Luco:1986, Bradford:2004} and all have found different natural frequencies and damping ratios within an expected range. As stated by Luco (\citeyear{Luco:1988}) “reduction in resonant frequencies mainly due changes in soil, foundation or superstructure remains an interesting question.”      

Furthermore, larger displacements from the most recent tests can also be seen on the estimated rigid body response with one key difference, the rigid body rocking behavior in the X-DOF and Y-DOF was larger in 2003 as seen in Figures 3.6(c) and 3.10(c) respectively. This difference can be attributed to the location of the applied force during such test which is further up next. 

\subsubsection{Potential Sources of Discrepancy}

The main differences between recent tests and those from 20 years ago were identified but are not limited to:

\begin{itemize}
    \item Condition of surrounding soil;
    \item Location and direction of applied forces;
    \item Magnitude of applied forces and scaling factors; and
    \item Noise in recorded data, data processing techniques and exact location of accelerometers.
\end{itemize}

First, during the 2003 tests, the east side next to the reaction mass was an empty pit backfilled after completion the test. This lack of surrounding soil would explain why the total X-DOF displacement of the top corner accelerometers was larger in 2003 which was not observed in any other group or any other degree of freedom. The dry climate conditions were similar during both studies, thus presence of water in the soil was not considered as a possible source of discrepancy.

Second, the direction of the applied forces was different in each case. The shakers from 2003 applied forces in the X or Y direction separately depending on the degree of freedom being tested. Meanwhile, in 2022 due to the V-shape configuration of the actuators, forces were applied simultaneously in the X and Y directions even though only one degree of freedom was tested at a time. Furthermore, none of the tests applied the forces at the center of gravity of the reaction mass, which led to an unintended torque exciting the rotational degrees of freedom. The shakers from 2003 were located at the top of the structure further away from the center of gravity, applying a larger moment when compared to that of 2022. This may explain why the rocking rigid body response was larger in the first tests, despite translation being greater in the most recent results. The simplified rigid model was once again used to analyze the influence of the location of applied forces, two cases were compared with loads applied at the top and closer to the center of gravity of the block respectively, while eliminating other variables such as force amplitude, scaling, and noise. The model showed the expected results that forces further away from the center of gravity generated larger displacements at the top and larger rotations. 

In addition, the magnitude of the applied forces, which determines the scaling factor, presented large differences between tests. As discussed in the test of linearity, the scaling factor plays an important role in the discrepancies since the linear scaling is only an approximation. Figure 4.3 shows a comparison of the applied forces and scaling factors used in the past and most recent tests.During the 2022 stepped frequency tests, the actuators maintained a relatively constant amplitude regardless of the excitation frequency which was determined via LSF. The force from the 2003 shakers on the other hand was a function of the frequency, which resulted in largely different scaling factors between the past and recent tests. Consequently, large scaling along with the non-linearity of the system introduced variability among results from different tests. 

\begin{figure}[H]
 \centering
 {\includegraphics[width=.7\linewidth]{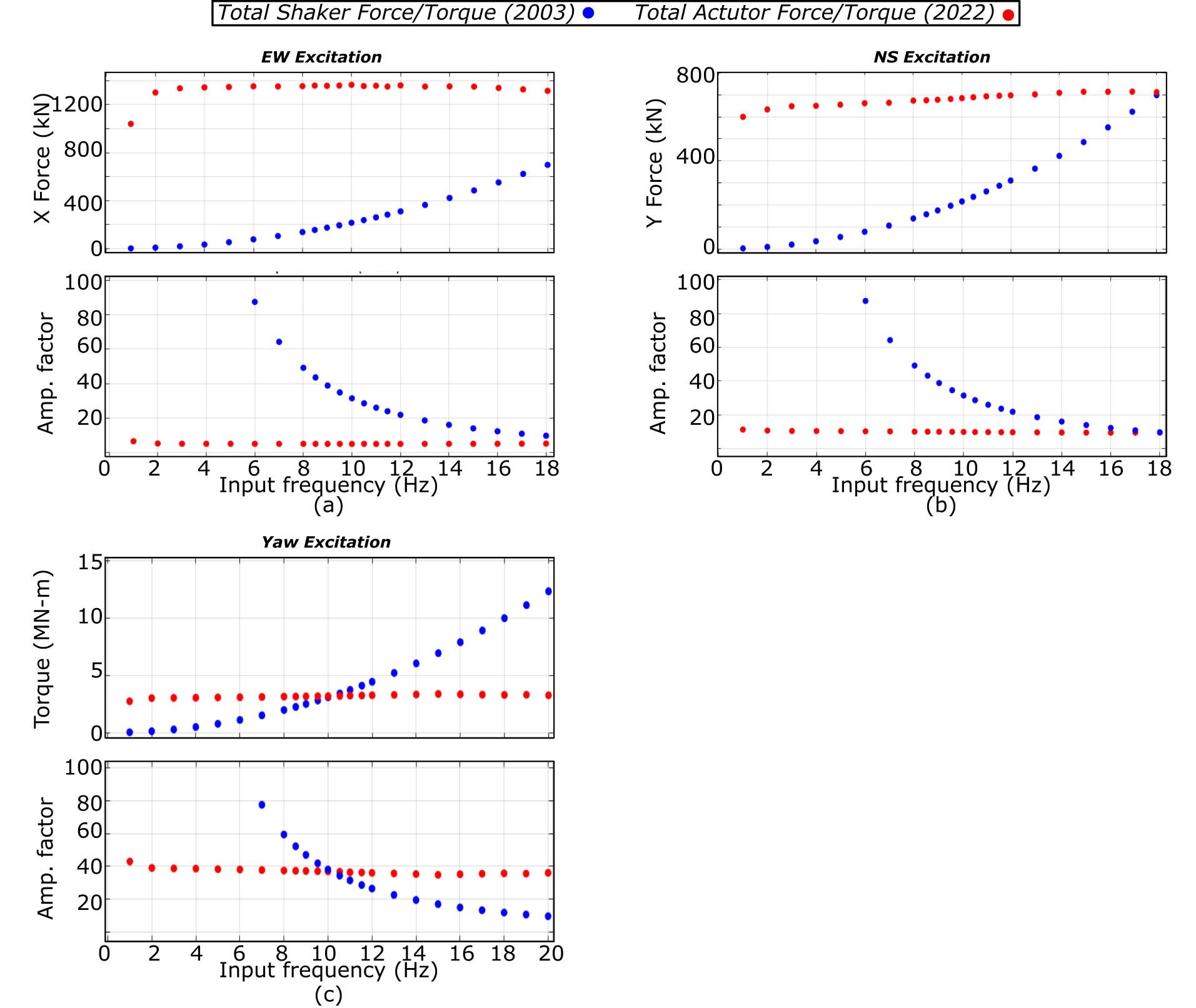} }
 \caption{Comparison of applied forces and scaling factors between tests from 2003 (blue) and 2022 (blue) under excitation in the (a) East-West, (b) North-South direction and (c) Yaw}
  \label{fig4.3}
\end{figure}

Moreover, data processing techniques added another layer or variability to the results, and even though it is complex to quantify its influence, it could not be left out of discussion. As mentioned in the description of the raw data, recorded response had the presence of super-harmonics that could not be filtered out, exciting the structure at more than one frequency. Additionally, the actuator forces applied in the X and Y direction simultaneously added noise to the response in the degree of freedom being tested. The noise in the data played a role when selecting the data filtering parameters such as the corner frequencies and the degree of the filter, which subsequently affect the computed displacements. 

Finally, differences in displacement were observed in the tenths of a millimeter. This might come up as a large relative difference considering a maximum absolute displacement close to 0.3mm, but a very small one when compared to the size of the structure. Therefore, the results can be confidently used as an estimation of the order of magnitude of the displacements if the maximum actuator forces were to be applied.

\subsection{Contribution of Rigid Body Motion}

Section 3 presented the percentage of total displacement that could be described with Rigid Body Motion. When analyzing the deformation pattern in all degrees of freedom it was observed that RBM cannot capture the vertical deformation of the bottom slab. In the case of the east-west and North-south response, vertical displacement came from rocking and not from slab deformation, thus, both translation and rocking were 90\% and 80\% RBM respectively for the East-west and north south excitation. The response of other two degrees of freedom, Yaw and vertical, on the other hand, presented a smaller contribution. In Figure 3.15, the applied moment induced vertical displacements at the top and bottom of the reaction mass that were entirely deformational not represented by RBM. Finally, RBM was able to represent part of the vertical displacement but not the additional displacements caused by the deformation of the slab. These results provided an insight of the limitations and the extent of the response that can be accurately represented with a model assumed rigid. 

Furthermore, the rigid body motion results in the East-west (X-DOF) and North-South (Y-DOF) indicated that the deformation took place mainly in the soil. Simultaneously, the test of linearity and comparison between tests provided the best agreement in these degrees of freedom as well. Meanwhile, the opposite was observed on the response to the Yaw excitation which had larger differences when comparing across tests. This response presented the least contribution of rigid body motion and the largest scaling factors. These results suggested that the soil behavior was linear at small displacements, and the linear scaling of results provided a better approximation when the deformation occurred mostly in the soil and not on the reaction mass. The linear behavior of the soil was congruent with the magnitude of the applied forces, which generated stresses of less than 1\% of the bearing capacity of the soil. 

The deformation pattern and percentages of rigid body motion presented in the results section and table 3.1 correspond to the response obtained under the frequency that produced the maximum displacement: 10Hz for X and Y-DOF and 14Hz and 15Hz for Yaw and Z-DOF, respectively. To have more clarity about the rigid body response at other frequencies, the average RBM contribution was computed for each of the frequencies tested and results are summarized in Figure 4.4. It was observed that for the translational X, Y, Z excitation, the contribution of rigid body motion was mostly constant with a variation of less than 10\%. The Yaw excitation, which had the least contribution of RBM, showed more variation depending on the frequency, confirming that this response cannot be appropriately captured with a rigid model. Interestingly it was observed that at higher frequencies past peak displacement, the RBM contribution kept decreasing. This behavior suggests that towards higher frequencies and smaller displacements, deformation occurs more on the reaction mass and less on the surrounding soil.  

\begin{figure}[H]
 \centering
 {\includegraphics[width=.8\linewidth]{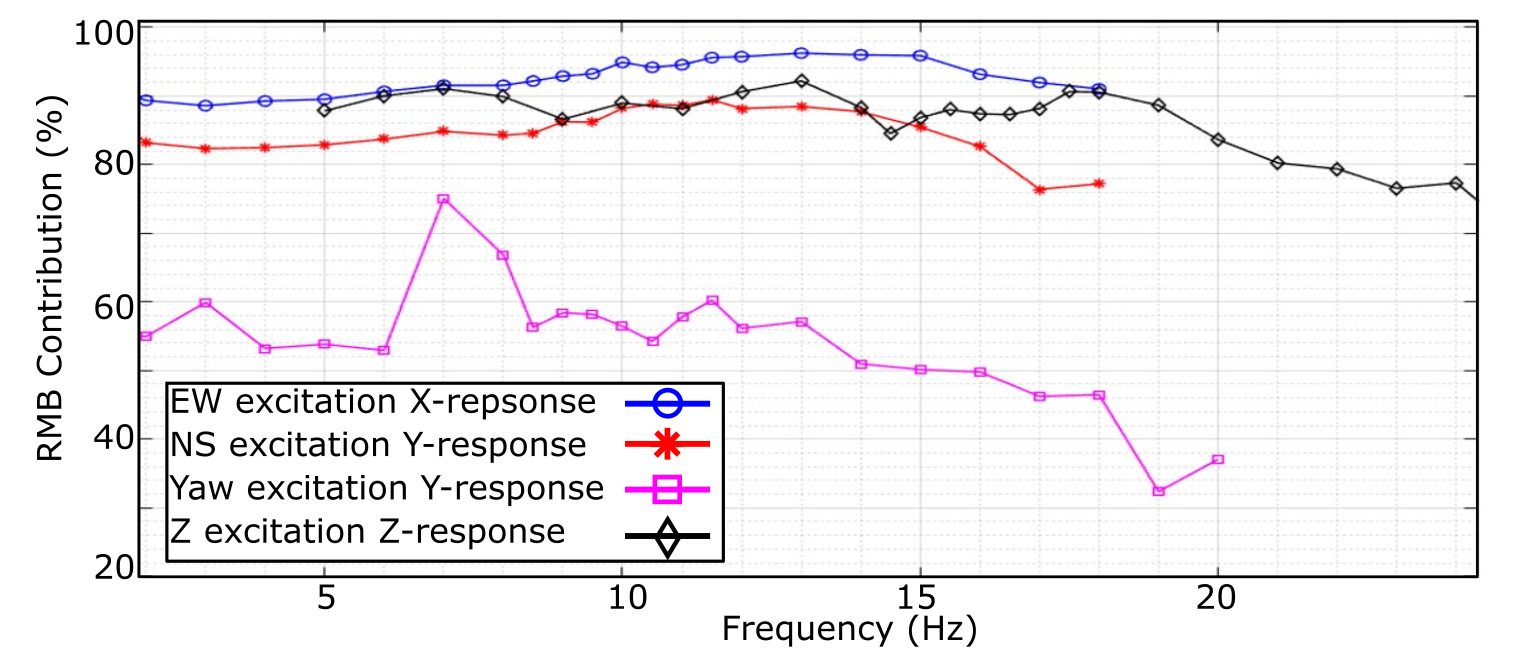} }
 \caption{Average Contribution of rigid body motion to total displacement for all tests at all frequencies tested.}
  \label{fig4.4}
\end{figure}

\section{CONCLUSIONS}

This report presented the results from forced vibration tests carried on the LHPOST’s reaction mass once the shake table upgrade had been completed. Displacement results were used to obtain the frequency response curves and deformation pattern of all the degrees of freedom tested, and to determine the natural frequencies, damping ratio and modes of vibration of the system. Despite the difficulties presented while performing the test, such as the unintended excitation of other degrees of freedom, non-constant force amplitude, and presence of super-harmonics in the excitation and response, fundamental frequencies of interest could still be identified. Furthermore, to provide more confidence on the computed displacement results, the data processing techniques were evaluated with synthetic data from a simplified 3D mass spring model.  

To obtain frequency response curves and compare displacement across tests, results were amplified to the same force amplitude. A test of linearity was carried to evaluate the scaling of results in a non-linear system. The test of linearity compared results from a set of tests where the only variable was the applied force. Comparison of frequency response curves showed a good agreement between tests indicating a fair degree of linearity in the system at small displacements. Test of linearity results from the X and Y degrees of freedom demonstrated that the linear scaling of results was a better approximation when the response was mainly in rigid body motion and deformation took place mostly on the soil. Furthermore, although the linearity was relative to only the most recent set of tests, results still provided confidence on the scaled displacements, denoting that the peak values presented in this report are a good approximation of the expected displacements if the actuators were to apply their maximum nominal force. 

Moreover, the results showed good agreement when compared to the study from 2003 validating the replicability of forced vibration tests. Even though the relative difference in displacement appeared large, they were only tenths of a millimeter compared to size of the reaction mass. After accounting for the sources of discrepancy between past and present tests, the good agreement in results suggested that the reaction mass-soil system did not present any major change of its dynamic properties once the upgrade was completed. 

A natural frequency of 9.5Hz was determined in the East-west direction based on the peak in the frequency response, and a vibration mode combining X-translation and pitch was observed. Response in the North-South direction presented its largest displacements between 10Hz and 12Hz without a clear peak. The deformation pattern presented rocking as well in a Y-translation-plus-roll mode of vibration. It should be noted that the displacement in these two degrees of freedom was mostly rigid body motion. Furthermore, a clear peak displacement was observed in the Yaw excitation frequency response, suggesting a natural of 14 Hz. This mode of vibration did not present any rocking, and it was mostly deformation at the top structure with very little contribution from rigid body motion. Finally, results from the vertical excitation suggest a natural frequency of 15Hz although the frequency response curve peaked was not so clear between 13Hz-16Hz. This mode of vibration had a considerable contribution from rigid body motion, mainly at the corners where the reaction mass is solid. Nonetheless, the bottom slab presented larger deformations that would be underestimated if only rigid body displacement is considered.

For future research, the data presented in this report will contribute towards the development of a computational model able to accurately represent the reaction mass-soil system. The observed dynamic response provided important information about the level of complexity required to properly represent the structure-soil system in all degrees of freedom. Finally, the experimental data will be a helpful resource to evaluate the current techniques used to model soil-structure interaction.

%%%%%%%%%%%%%%%%%%%%%%%%%%%

\bibliographystyle{apalike}
\bibliography{2references}

\end{document}